\documentclass[journal]{IEEEtran}
\usepackage[T1]{fontenc}
\usepackage{cite}
\usepackage{amsmath,amssymb,amsfonts}
\usepackage{algorithm}
\usepackage{algorithmic}
\usepackage{graphicx}
\usepackage{textcomp}
\usepackage{xcolor}
\usepackage{subfigure}
\usepackage{booktabs}
\usepackage{multirow}
\usepackage{diagbox}
\usepackage{url}
\usepackage{threeparttable}
\usepackage{array}
\usepackage{textcase}
\usepackage{rotating}
\usepackage{etoolbox}
\makeatletter
\def\UrlAlphabet{%
      \do\a\do\b\do\c\do\d\do\e\do\f\do\g\do\h\do\i\do\j%
      \do\k\do\l\do\m\do\n\do\o\do\p\do\q\do\r\do\s\do\t%
      \do\u\do\v\do\w\do\x\do\y\do\z\do\A\do\B\do\C\do\D%
      \do\E\do\F\do\G\do\H\do\I\do\J\do\K\do\L\do\M\do\N%
      \do\O\do\P\do\Q\do\R\do\S\do\T\do\U\do\V\do\W\do\X%
      \do\Y\do\Z}
\def\UrlDigits{\do\1\do\2\do\3\do\4\do\5\do\6\do\7\do\8\do\9\do\0}
\g@addto@macro{\UrlBreaks}{\UrlOrds}
\g@addto@macro{\UrlBreaks}{\UrlAlphabet}
\g@addto@macro{\UrlBreaks}{\UrlDigits}

\patchcmd{\@algocf@start}
  {-1.5em}
  {0pt}
  {}{}

\makeatother

\newcommand{\Times}[2]{${\text{#1}\times\text{#2}}$}
\newcommand{\SNR}[2]{${\text{SNR} #1 \text{#2}\;\text{dB}}$}
\def\post{\textit{a posteriori }}

\graphicspath{{figures/}}

\def\nt{N_{\rm t}}
\def\nr{N_{\rm r}}

\begin{document}
\ifdefined \GramaCheck
  \newcommand{\CheckRmv}[1]{}
  \newcommand{\figref}[1]{Figure 1}%
  \newcommand{\tabref}[1]{Table 1}%
  \newcommand{\secref}[1]{Section 1}
  \newcommand{\algref}[1]{Algorithm 1}
  \renewcommand{\eqref}[1]{Equation 1}
\else
  \newcommand{\CheckRmv}[1]{#1}
  \newcommand{\figref}[1]{Fig.~\ref{#1}}%
  \newcommand{\tabref}[1]{Table~\ref{#1}}%
  \newcommand{\secref}[1]{Sec.~\ref{#1}}
  \newcommand{\algref}[1]{Algorithm~\ref{#1}}
  \renewcommand{\eqref}[1]{(\ref{#1})}
\fi
\newtheorem{theorem}{Theorem}
\newtheorem{proposition}{Proposition}
\newtheorem{assumption}{Assumption}
\newtheorem{definition}{Definition}
\newtheorem{condition}{Condition}
\newtheorem{property}{Property}
\newtheorem{remark}{Remark}
\newtheorem{lemma}{Lemma}
\newtheorem{corollary}{Corollary}
%
\title{Near-Optimal MIMO Detection Using Gradient-Based MCMC in Discrete Spaces}
%

%
%
\author{Xingyu~Zhou,~\IEEEmembership{Graduate Student Member,~IEEE,}
      Le~Liang,~\IEEEmembership{Member,~IEEE,}
      Jing~Zhang,~\IEEEmembership{Member,~IEEE,}
      Chao-Kai~Wen,~\IEEEmembership{Fellow,~IEEE,}
      and~Shi~Jin,~\IEEEmembership{Fellow,~IEEE}
\thanks{X.~Zhou, L.~Liang, J.~Zhang, and S.~Jin are with the National Mobile
Communications Research Laboratory, Southeast University, Nanjing 210096, China
(e-mail: \protect \url{xy_zhou@seu.edu.cn}; lliang@seu.edu.cn; jingzhang@seu.edu.cn; jinshi@seu.edu.cn). L. Liang is also with the Purple Mountain Laboratories, Nanjing 211111, China.}
\thanks{C.-K.~Wen is with the Institute of Communications Engineering,
National Sun Yat-sen University, Kaohsiung 80424, Taiwan
(e-mail: chaokai.wen@mail.nsysu.edu.tw).}
}

%
%

\maketitle

\begin{abstract}
The discrete nature of transmitted symbols poses challenges for achieving optimal detection in multiple-input multiple-output (MIMO) systems associated with a large number of antennas.
Recently, the combination of two powerful machine learning methods, Markov chain Monte Carlo (MCMC) sampling and gradient descent, has emerged as a highly efficient solution to address this issue. 
However, existing gradient-based MCMC detectors are heuristically designed and thus are theoretically untenable.  
To bridge this gap, we introduce a novel sampling algorithm tailored for discrete spaces. This algorithm leverages gradients from the underlying continuous spaces for acceleration while maintaining the validity of probabilistic sampling. 
We prove the convergence of this method and also analyze its convergence rate using both MCMC theory and empirical diagnostics. 
On this basis, we develop a MIMO detector that precisely samples from the target discrete distribution and generates posterior Bayesian estimates using these samples, whose performance is thereby theoretically guaranteed. 
Furthermore, our proposed detector is highly parallelizable and scalable to large MIMO dimensions, positioning it as a compelling candidate for next-generation wireless networks. 
Simulation results show that our detector achieves near-optimal performance, significantly outperforms state-of-the-art baselines, and showcases resilience to various system setups.

\end{abstract}


%
\IEEEpeerreviewmaketitle

\section{Introduction}

Over the past two decades, the multiple-input multiple-output (MIMO) technology has played a crucial role in improving spectral efficiency and network throughput of wireless communication systems \cite{bjornson2017massive}.
To enhance data rates and coverage even further, future generations of wireless networks are expected to witness an unprecedented rise in the number of antennas, leading to  extra-large-scale MIMO systems \cite{bjornsonMassiveMIMOReality2019,zhang2020prospective}. However, this dramatically increased number of antennas results in the curse of dimensionality in symbol detection---a significant roadblock to realizing the full potential of MIMO.

The optimal maximum \post (MAP) detection involves exhaustive enumeration and is tractable only for the most trivial cases \cite{yangFiftyYearsMIMO2015}.
Sphere decoding (SD) has been introduced to reduce the complexity of MAP detection while  approaching near-optimal performance \cite{hochwaldAchievingNearcapacityMultipleantenna2003,guoAlgorithmImplementationKbest2006}. Nonetheless, the substantial tree searches still entail complexity that exponentially grows with the system dimension. Linear detectors, such as the linear minimum mean square error (MMSE) method, are widely acknowledged for their low complexity. However, they are prone to  severe performance degradation in high-order modulation or correlated channels.

Posterior inference problems like MIMO detection generally involve multidimensional function integration or maximization  and become challenging as the dimension increases.
Recently, stochastic sampling techniques \cite{bishopPatternRecognitionMachine}, also known as Monte Carlo methods, have shown great potential in addressing the curse of dimensionality in posterior inference.  
These methods sample from certain probability distributions of interest and generate reliable estimates based on the drawn samples. 
Among them, the Markov chain Monte Carlo (MCMC) method \cite{brooks2011handbook} has been popular and has found extensive applications in various communication signal processing tasks that can be regarded as posterior inferences \cite{doucetMonteCarloMethods2005,chenConvergenceAnalysesComparisons2002}. 
In MCMC, statistical inferences are developed by simulating a Markov chain to generate a sequence of samples that converge to the target posterior distribution and performing Monte Carlo summation using the converged samples. 
This approach allows for the reduction of the exponential complexity associated with the dimension to a polynomial level.
In particular, for MIMO detection, the MCMC method is highly acclaimed for its hardware-friendly architecture \cite{larawayImplementationMarkovChain2009} and attainment of remarkable performance with relatively low complexity \cite{farhang-boroujenyMarkovChainMonte2006,hedstromAchievingMAPPerformance2017}.

Discrete data is ubiquitous in real-world applications, ranging from genomes in biology and texts in linguistics to bits and symbols in wireless communications.
Exact inference on this type of data, such as optimal MIMO detection, requires sampling from discrete distributions, which has long been recognized as more challenging compared to sampling from continuous distributions  \cite{grathwohlOopsTookGradient2021}.
Gibbs sampling \cite{geman1984stochastic}, a classical MCMC method, is well known as a generic solution to this task; however, this method is subject to the \textit{sequential} update of variables and hence suffers from low efficiency when the dimension is large or the distribution is highly correlated among its variables. 
As data dimensions scale up, developing efficient sampling algorithms for discrete distributions is urgent. 

In recent years, MCMC has been combined with the gradient descent method, formulating a promising machine learning framework for nonconvex optimization \cite{maSamplingCanBe2019}. This gradient-based MCMC paradigm was initially developed for continuous spaces and has demonstrated enhanced efficiency than conventional sampling and optimization methods in solving inference problems.  
Owing to its superiority in continuous spaces, new research trends have revolved around generalizing gradient-based MCMC to discrete spaces. 
The core concept is to leverage gradients of the underlying continuous function to navigate the sampling toward the target discrete distribution. 
Based on this idea, gradients have been used to accelerate Gibbs sampling by indicating the variables that should be prioritized for updating \cite{grathwohlOopsTookGradient2021}. 
To further improve sampling efficiency, the Langevin algorithm \cite{robertsExponentialConvergenceLangevin1996,robertsLangevinDiffusionsMetropolisHastings2002}, which is an advanced gradient-based MCMC method originally developed for continuous spaces, has been extended to discrete spaces in \cite{zhangLangevinlikeSamplerDiscrete2022,rhodesEnhancedGradientbasedMCMC2022}. Unlike the Gibbs sampler, the Langevin method updates all variables in \textit{parallel} to enable large movements and utilizes gradients to direct the sampling to high probability regions of the target distribution, thus showcasing orders-of-magnitude improvements in efficiency.

Given that MIMO detection is a typical posterior inference problem within the discrete space of the transmitted symbols, it is intuitively appealing to apply advanced gradient-based MCMC methods that have been developed for discrete spaces. 
In fact, this application has been extensively explored in previous works such as \cite{gowdaMetropolisHastingsRandomWalk2021,zhou2023gradient,zilbersteinAnnealedLangevinDynamics2022a,wuStochasticGradientLangevin2022}, achieving both exceptional performance and high efficiency.  
Specifically, Newton's method \cite{Luenberger1973introduction} was leveraged to accelerate MCMC sampling in \cite{gowdaMetropolisHastingsRandomWalk2021}. 
In their proposed detector, MCMC's exploration was conducted along the preconditioned gradient descent direction in the continuous-relaxed space, followed by a Metropolis-Hastings (MH) correction \cite{hastingsMonteCarloSampling1970}. Nevertheless, the derivation of the correction step was based on heuristics. 
In \cite{zhou2023gradient}, additional improvements were investigated by employing  Nesterov's accelerated gradient method \cite{nesterovMethodSolvingConvex1983} to expedite MCMC sampling. This method avoided the computationally intensive matrix inversion required by Newton's method and hence enhanced the scalability of the detector. Moreover, an annealed Langevin algorithm was developed in \cite{zilbersteinAnnealedLangevinDynamics2022a} by setting multiple noise levels with decreasing variances to mimic the discrete prior of the transmitted symbols. This approach enabled the computation of gradients and the transformation of the intricate discrete approximation into a simplified continuous approximation. Additionally, the Langevin-based MIMO detector was independently investigated in \cite{wuStochasticGradientLangevin2022}. However, their proposed detector did not consider the discrete nature of symbols and is limited to a specific modulation scheme.

Despite the rapid development, a notable deficiency of existing gradient-based MCMC detectors \cite{gowdaMetropolisHastingsRandomWalk2021,zhou2023gradient,zilbersteinAnnealedLangevinDynamics2022a,wuStochasticGradientLangevin2022} is their lack of theoretical guarantees. These schemes rely on heuristic designs due to the challenges posed by exact sampling in discrete spaces \cite{grathwohlOopsTookGradient2021,zhangLangevinlikeSamplerDiscrete2022} and give no convergence guarantees. More importantly, as shown in this paper, these heuristics can undermine the reliability of MCMC by generating samples that deviate from the target distribution for MIMO detection. Such divergence can lead to substantial errors in inference and severe performance degradation. Hence, the development of a gradient-based MCMC method that ensures precise sampling from the target discrete distribution holds great significance.

In this paper, we propose a near-optimal MIMO detector based on gradient-based MCMC that exactly samples in discrete spaces, ensuring the validity of stochastic sampling.
The proposed detector offers an efficient solution for computing posterior Bayesian estimates via the generation of important samples and the subsequent Monte Carlo summation. The performance is guaranteed by the convergence of the sampling algorithm and the law of large numbers behind Monte Carlo methods. We perform extensive numerical experiments to verify the effectiveness of our proposed detector. Results show that our method achieves near-optimal performance with a limited number of samples and significantly outperforms state-of-the-art detectors.\footnote{{The term ``near-optimal'' in this context refers to the proposed detector's ability, theoretically guaranteed, to approach optimal detection performance given sufficient computational resources. Importantly, the proposed detector offers significant complexity savings compared to the optimal MAP detector, rendering it a practically viable solution for MIMO systems with a large number of antennas.}}  Moreover, our proposed detector exhibits remarkable robustness to various channel environments and resilience to imperfect channel state information (CSI).
Additionally, our detector is scalable to large MIMO dimensions and highly parallelizable, making it a promising solution for extra-large-scale MIMO systems in the next-generation wireless networks. 
We summarize the contributions of this paper as follows.

\begin{itemize}
  \item \textbf{Gradient-Based MCMC Sampling for Discrete Distributions}: We have developed a discrete analog to the Metropolis adjusted Langevin algorithm (MALA) \cite{robertsLangevinDiffusionsMetropolisHastings2002} for the target discrete distributions within the context of the MIMO detection problem.
  This novel algorithm, referred to as DMALA, achieves high-quality sampling from discrete distributions by leveraging gradients from the underlying continuous function for acceleration, while strictly adhering to the systematic steps of MCMC.
  
  \item \textbf{Convergence Proof and Analysis}: We provide a theoretical proof of DMALA's convergence and analyze its convergence rate.
  This theoretically guaranteed property distinguishes our proposed algorithm from heuristic sampling algorithms used in existing gradient-based MCMC detectors.
  Additionally, we offer empirical diagnostics of the convergence (rate) to validate the theoretical findings. 
  
  \item \textbf{Achievement of Near-Optimal Detection}: Utilizing DMALA, we propose a MIMO detector that initially samples from the target discrete distribution and then computes Bayesian estimates (soft decisions) using the converged samples. The method employed for computing soft decisions is meticulously designed in alignment with Monte Carlo theory, setting it apart from conventional heuristic approaches. Consequently, the performance of our proposed detector is theoretically underpinned by the convergence of DMALA and the law of large numbers inherent in Monte Carlo methods. This near-optimal performance has been further corroborated through extensive numerical studies.
  
\end{itemize}

\textit{Notations:} Lowercase and uppercase boldface letters denote column vectors and matrices, respectively.
$\mathbf{A}^{-1}$ and $\mathbf{A}^{T}$ represent the inverse and transpose of a matrix $\mathbf{A}$, respectively.
$\mathbf{I}_N$ represents an $N\times N$ identity matrix. 
$\mathbb{R}$ is the set of real numbers.
$\mathbb{E}[\cdot]$ denotes the expectation operation. 
$\mathcal{U}(a,b)$ denotes a uniform distribution between $[a,b]$. 
$\mathcal{N}(\mu,\sigma^2)$ indicates a real-valued Gaussian distribution with mean $\mu$ and variance $\sigma^2$. 
$\|\cdot\|_F$ denotes Frobenius norm, and $\|\cdot\|$ denotes $l_2$ norm.
$|\cdot|$ represents the cardinality of a set.

\section{System Model and Preliminaries}

This section commences with an introduction to the system model of the MIMO detection problem. It then proceeds to provide an overview of the basic concepts underlying MCMC.

\subsection{System Model}  

Consider a MIMO system with $N_{\rm t}$ antennas for transmitting data streams and $N_{\rm r}$ antennas for receiving. The message bits are encoded and interleaved to generate the transmitted codeword, which is then partitioned into bit vectors. 
Each bit vector $\mathbf{b}\in\{\pm 1\}^{N_{\rm b}}$ is mapped into quadrature amplitude modulation (QAM) symbols with unit power in average, constituting the transmitted vector. The equivalent real-valued model for the MIMO transmission is given by  
\CheckRmv{
  \begin{equation}
    \mathbf{y} = \mathbf{Hx} + \mathbf{n},
    \label{eq:sys_model}
  \end{equation}
}  
where $\mathbf{x} \in \mathcal{A}^{N \times 1}$ denotes the equivalent real-valued symbol vector, where $N=2N_{\rm t}$ and $\mathcal{A}$ is the finite set of real-valued transmitted symbols with a cardinality of $|\mathcal{A}| = Q$. 
Therefore, we have $N_{\rm b} = N\log_2Q$.\footnote{{We assume that all elements of $\mathbf{x}$ are selected from a common finite set $\mathcal{A}$, implying the use of a single modulation scheme. However, extending the proposed method to handle cases where elements of $\mathbf{x}$ are drawn from different discrete spaces is straightforward.}}
Moreover, in \eqref{eq:sys_model}, $\mathbf{y} \in \mathbb{R}^{M \times 1}$ is the received real-valued signal with $M=2N_{\rm r}$, $\mathbf{H} \in \mathbb{R}^{M \times N}$ is the real-valued channel matrix, and $\mathbf{n} \in \mathbb{R}^{M \times 1}$ is the noise vector whose elements independently follow $\mathcal{N}(0,\sigma^2/2)$, where $\sigma^2/2$ is the noise variance per real element.

Given the observation $\mathbf{y}$ and assuming that the channel $\mathbf{H}$ is known, the posterior distribution of $\mathbf{x}$ is given by the discrete distribution $\pi(\mathbf{x}) = p(\mathbf{x}|\mathbf{y}) \propto p(\mathbf{x})p(\mathbf{y}|\mathbf{x})$, where $p(\mathbf{x})$ is the prior distribution, and $p(\mathbf{y}|\mathbf{x})$ is the likelihood.
When the prior distribution is uniform, the posterior distribution can be further expressed as
\CheckRmv{
  \begin{align}
    \pi(\mathbf{x}) =\frac{1}{Z} \exp \big(f(\mathbf{x})\big) \prod_{n=1}^N \mathbb{I}_{x_n \in \mathcal{A}}, 
    \label{eq:post}
  \end{align}
}
where $f(\mathbf{x})$ is a metric function given by 
\CheckRmv{
  \begin{align}
    f(\mathbf{x}) &= -\frac{1}{\sigma^2}\|\mathbf{y} - \mathbf{Hx}\|^2, \label{eq:f} 
  \end{align}
}
$Z$ is a normalization constant whose computation involves multidimensional summation and is generally intractable, $x_n$ denotes the $n$-th entry of $\mathbf{x}$, 
and $\mathbb{I}_{x_n \in \mathcal{A}}$ is an indicator function that takes the value of one only if $x_n \in \mathcal{A}$ and zero otherwise. 

The MIMO detector forwards soft outputs in terms of log-likelihood ratios (LLRs) to the subsequent channel decoder.
The optimal MAP detector computes the posterior LLR for the $k$-th element $b_k$ of $\mathbf{b}$ as  
\CheckRmv{
  \begin{equation}
    L_k = \log \frac{p(b_k=+1 | \mathbf{y})}{p(b_k=-1 | \mathbf{y})} = \log \frac{\sum_{\mathbf{x}\in \mathcal{A}^{N\times 1}_{k+}} \exp\big(f(\mathbf{x})\big)}{\sum_{\mathbf{x}\in \mathcal{A}^{N\times 1}_{k-}} \exp\big(f(\mathbf{x})\big)},
    \label{eq:exact_llr}
  \end{equation}
}
where $\mathcal{A}_{k+}^{N \times 1}$ and $\mathcal{A}_{k-}^{N \times 1}$ denote the subsets of $\mathcal{A}^{N\times 1}$ for the transmitted vector $\mathbf{x}$ mapped from $\mathbf{b}$, where $b_k$ corresponds to $+1$ and $-1$, respectively.
This computation involves evaluating two \post probabilities (APPs) that both require iteration over all possible values of $\mathbf{b}_{-k} = [b_1,\ldots,b_{k-1},b_{k+1},\ldots,b_{N_{\rm b}}]^T$ and the summation of $2^{N_{\rm b}-1}$ terms, causing prohibitive complexity for large $N$ and/or $Q$. 
Therefore, a low-complexity alternative should be developed.

\subsection{Basics of MCMC} \label{sec:basics}

We begin with the basics of Monte Carlo methods. Let $X$ denote a discrete random variable, possibly multidimensional, with a distribution $g(X)$. Consider evaluating the expectation of some function $h(X)$ with respect to $g(X)$, i.e., 
\CheckRmv{
  \begin{equation}
    \mathbb{E}_{g}[h(X)]=\sum_{x\in \mathcal{X}} h({x})g(x),
    \label{eq:expectation}
  \end{equation}
}
where $\mathcal{X}$ is the domain of $X$. Suppose that a set of $S$ samples $\{x^{[s]}\}_{s=1}^{S}$ with $x^{[s]} \sim g(X)$ is available, then the empirical average   
\CheckRmv{
  \begin{equation}
    \bar{h} = \frac{1}{S} \sum_{s=1}^{S}  h\left({x}^{[s]}\right)
    \label{eq:mc}
  \end{equation}
}
is an unbiased estimate of the expectation $\mathbb{E}_{g}[h(X)]$, i.e., $\bar{h}\stackrel{\text { a.s. }}{\longrightarrow}\mathbb{E}_{g}[h(X)]$ as $S\to \infty$. A notable feature of this approach is that the required number of samples $S$ to achieve acceptable accuracy is weakly dependent on the dimension of $X$ \cite{farhang-boroujenyMarkovChainMonte2006}. Therefore, the exponential complexity that is often encountered when computing the summation in \eqref{eq:expectation} can be mitigated.

Importance sampling (IS) is a classical Monte Carlo method that has the potential to further reduce the variance of $\bar{h}$ \cite{farhang-boroujenyMarkovChainMonte2006}. 
IS introduces an auxiliary distribution $g_{\rm a}(X)$ on $\mathcal{X}$, {which has the same support as $g(X)$,} and approximates the expectation by
\CheckRmv{
  \begin{equation}
    \frac{1}{S} \sum_{s=1}^{S} \frac{g(x^{[s]})}{g_{\rm a}(x^{[s]})} h\left(x^{[s]}\right), \quad x^{[s]} \sim g_{\rm a}(X).
    \label{eq:is_mc}
  \end{equation}
}
It turns out that by a wise selection of $g_{\rm a}(X)$, an accurate approximation can be obtained using fewer samples than the scheme in \eqref{eq:mc} \cite{bishopPatternRecognitionMachine,kashifMonteCarloEqualization2008}.

The above Monte Carlo methods require samples from a target distribution. 
MCMC is a popular and generic method for realizing this aim, effective for a wide range of distributions, and scales well with the dimension of the space \cite{bishopPatternRecognitionMachine}. Specifically, the MCMC method simulates a Markov chain $x^{(1)}, x^{(2)}, \ldots,x^{(t)},\ldots$ that is regulated by a transition kernel $P(\cdot|\cdot)$, where $t$ is the time step index. Each state in the chain corresponds to a sample, and given the current state $x^{(t)}$, the new state is generated as $x^{(t+1)}\sim P(\cdot|x^{(t)})$. The transition kernel is designed to ensure that the chain's stationary distribution coincides with the target distribution, e.g., $g$ or $g_{\rm a}$. In this manner, the samples $\{x^{(t)}\}$ asymptotically converge to the target distribution for large $t$. These converged samples can then be used for the Monte Carlo summation in \eqref{eq:mc} or \eqref{eq:is_mc}. As the most popular MCMC-type scheme, the MH algorithm first samples a simple proposal distribution $x^{\prime}\sim q(\cdot|x^{(t)})$. 
To ensure convergence to the target distribution, such as $g=\pi$ without loss of generality, the proposal $x^{\prime}$ is then accepted as the new state $x^{(t+1)}$ with a probability \cite{hastingsMonteCarloSampling1970}
\CheckRmv{
\begin{equation}
  \min  \left\{1, \frac{\pi({x}^{\prime})q({x}^{(t)}|{x}^{\prime})}{\pi({x}^{(t)})q({x}^{\prime}|{x}^{(t)})}\right\}.
  \label{eq:acc_prob_general}
\end{equation}
}
Otherwise, the current state is retained as the new state, i.e., $x^{(t+1)}=x^{(t)}$.
With this criterion, the corresponding Markov chain admits $\pi$ as the stationary distribution \cite{doucetMonteCarloMethods2005}.
Meanwhile, this algorithm eliminates the need to evaluate the normalization constant of $\pi$ since $\pi$ appears as a ratio in \eqref{eq:acc_prob_general}.

\section{DMALA-Based MIMO Detection} \label{sec:dmala}
In this section, we derive the DMALA-based MIMO detector. Initially, we develop a highly efficient gradient-based MCMC sampling algorithm tailored to discrete spaces. Subsequently, we establish the convergence of the proposed sampling algorithm and substantiate our claims through empirical verification.
Then, we discuss the utilization of the samples for soft decisions, specifically in terms of LLR computation, and examine the computational complexity of our proposed detector.
Overall, the proposed detector achieves high accuracy in approximating the exact posterior Bayesian estimate in \eqref{eq:exact_llr}, while mitigating the prohibitive complexity.

\subsection{Gradient-Based MCMC Sampling in Discrete Spaces}  \label{sec:algorithm}

For ease of exposition, we target the posterior distribution $\pi(\mathbf{x})$ in \eqref{eq:post} for sample generation. It is noteworthy that this probability mass function can be regarded as a restriction of a continuous distribution defined over the domain $\mathbb{R}^{N\times 1}$ to the discrete subset $\mathcal{A}^{N\times1}$. Therefore, gradients from the \textit{logarithm of the underlying continuous distribution}, i.e., $f(\mathbf{x})$ in \eqref{eq:f}, are informative for the sampling of the discrete distribution.\footnote{{The proposed algorithm and the subsequent theoretical analysis are not limited to the distribution in \eqref{eq:post} and can be generalized to different $f(\mathbf{x})$. Moreover, we consider gradients from the log-probability to simplify calculation.}} 

We start with introducing the Langevin algorithm \cite{robertsExponentialConvergenceLangevin1996,robertsLangevinDiffusionsMetropolisHastings2002}, which is a powerful gradient-based MCMC method in continuous spaces. Initialized with $\mathbf{x}^{(1)} \in \mathbb{R}^{N \times 1}$, each sampling iteration of this algorithm generates a proposal vector \cite{robertsExponentialConvergenceLangevin1996,robertsLangevinDiffusionsMetropolisHastings2002,wellingBayesianLearningStochastic} 
\CheckRmv{
  \begin{equation}
    \mathbf{x}^{\prime} = \mathbf{x}^{(t)} + \frac{\alpha}{2} \nabla f(\mathbf{x}^{(t)}) + \sqrt{\alpha} \mathbf{w}^{(t)},  
    \label{eq:langevin}
  \end{equation}
}
where $t$ is the sampling iteration index (equivalent to the time step index of the underlying Markov chain), $\alpha > 0$ is the step size, and $\mathbf{w}^{(t)}$ is a random perturbation that follows $\mathcal{N}(\mathbf{0}, \mathbf{I}_N)$, where $\mathbf{0}$ is a zero vector. $\nabla f$ is the gradient of $f(\mathbf{x})$ given by
\CheckRmv{
  \begin{equation}
    \nabla f(\mathbf{x}) = \frac{2}{\sigma^2} \mathbf{H}^T (\mathbf{y} - \mathbf{Hx}).
    \label{eq:grad} 
  \end{equation}
} 
This gradient facilitates efficient exploration of high probability regions of the sample space. 
It should be noted that the update rule in \eqref{eq:langevin} can be viewed as drawing the proposal vector from the Gaussian distribution 
\CheckRmv{
  \begin{equation}
     \mathcal{N}\left(\mathbf{x}^{(t)} + \frac{\alpha}{2} \nabla f(\mathbf{x}^{(t)}),  \alpha \mathbf{I}_N\right).
    \label{eq:proposal_cont}
  \end{equation} 
}

Based on \eqref{eq:langevin} and considering the variable domain $\mathcal{A}^{N\times 1}$, we derive the discrete proposal function 
\CheckRmv{
  \begin{align}
    q&(\mathbf{x}^{\prime} | \mathbf{x}^{(t)}) \nonumber \\ 
    &= \frac{\exp\left(-\frac{1}{2\alpha}\|\mathbf{x}^{\prime}-\mathbf{x}^{(t)} - \frac{\alpha}{2} \nabla f(\mathbf{x}^{(t)}) \|^2\right)}{Z_{\mathcal{A}}(\mathbf{x}^{(t)})}\prod_{n=1}^N \mathbb{I}_{x_n^{\prime} \in \mathcal{A}},
    \label{eq:dis_prop}
  \end{align} 
}
where $x_n^{\prime}$ is the $n$-th element of $\mathbf{x}^{\prime}$, and the normalization constant $Z_{\mathcal{A}}(\mathbf{x}^{(t)})$ is given by 
\CheckRmv{
  \begin{equation}
    Z_{\mathcal{A}}(\mathbf{x}^{(t)}) = \sum_{\mathbf{x}^{\prime} \in \mathcal{A}^{N\times 1}} \exp \Big(-\frac{1}{2\alpha}\|\mathbf{x}^{\prime}-\mathbf{x}^{(t)} - \frac{\alpha}{2} \nabla f(\mathbf{x}^{(t)}) \|^2\Big),
  \end{equation}
}
whose computation is generally intractable since it requires traversal over the full space of size $Q^{N}$.
However, a distinct feature of the proposal in \eqref{eq:dis_prop} is that it enjoys an \textit{elementwise} factorization as \cite{zhangLangevinlikeSamplerDiscrete2022}:
\CheckRmv{
  \begin{equation}
    q(\mathbf{x}^{\prime} | \mathbf{x}^{(t)}) = \prod_{n=1}^{N} q_n(x^{\prime}_n | x^{(t)}_n),
    \label{eq:dis_prop_fac1}
  \end{equation}
} 
where $q_n(x^{\prime}_n | x^{(t)}_n)$ is a categorical distribution given by
\CheckRmv{
  \begin{align}
    q_n(x^{\prime}_n | x^{(t)}_n) &= \varsigma \big(\mu(x^{\prime}_n)\big)\mathbb{I}_{x_n^{\prime} \in \mathcal{A}}, \label{eq:dis_prop_fac2}
  \end{align}
}
where $\mu(\cdot)$ denotes one term in the factorization of the $l_2$ norm in \eqref{eq:dis_prop}, and $\varsigma(\cdot)$ is a softmax function. These two functions are given by
\CheckRmv{
  \begin{align}
    \mu(x^{\prime}_n) 
    &= \frac{1}{2}\left[\nabla f(\mathbf{x}^{(t)})\right]_n (x^{\prime}_n - x^{(t)}_n) - \frac{(x^{\prime}_n - x^{(t)}_n)^2}{2 \alpha} \label{eq:exp_fac} 
  \end{align}
}
and
\CheckRmv{
  \begin{equation}
    \varsigma\big(\mu(x^{\prime}_n)\big) = \frac{\exp\big(\mu(x^{\prime}_n)\big)}{\sum_{x_n^{\prime}\in \mathcal{A}} \exp\big(\mu(x^{\prime}_n)\big)}, \label{eq:softmax}
  \end{equation}
}
where $\left[\nabla f(\mathbf{x}^{(t)})\right]_n$ is the $n$-th element of the gradient vector $\nabla f(\mathbf{x}^{(t)})$. This factorization enables the parallel update of each element in $\mathbf{x}^{(t)}$, i.e., ${x^{\prime}_n \sim q_n(\cdot | x^{(t)}_{n})}, n=1,\ldots,N$, after gradient computation, whose cost is only $\mathcal{O}(N^2)$. Therefore, the overall computational cost of constructing the proposal in \eqref{eq:dis_prop_fac2} and parallel updating scales  polynomially, rather than exponentially, with $N$. 

The update rule in \eqref{eq:langevin} originates from discretizing the stochastic differential equation of Langevin diffusion \cite{robertsExponentialConvergenceLangevin1996}. Due to discretization errors, the unadjusted Langevin algorithm, i.e., directly letting $\mathbf{x}^{(t+1)} = \mathbf{x}^{\prime}$, typically suffers from asymptotic bias towards the target distribution \cite{robertsLangevinDiffusionsMetropolisHastings2002,zhangLangevinlikeSamplerDiscrete2022,wellingBayesianLearningStochastic}.
To deal with this issue, we integrate the MH adjustment with the discrete Langevin proposal \eqref{eq:dis_prop_fac2} to ensure the reversibility of the Markov chain $\{\mathbf{x}^{(t)}\}$ and the convergence to the target distribution \cite{hastingsMonteCarloSampling1970}, resulting in \textit{DMALA}.
Specifically, after generating $\mathbf{x}^{\prime}$ using the proposal distribution in \eqref{eq:dis_prop_fac2}, the MH adjustment introduced in \secref{sec:basics} is performed to
accept $\mathbf{x}^{\prime}$ with a probability given by
\CheckRmv{
  \begin{align}
    A(\mathbf{x}^{\prime}|\mathbf{x}^{(t)}) 
    &= \min \left\{1, \exp\big(f(\mathbf{x}^{\prime}) - f(\mathbf{x}^{(t)})\big) \frac{q(\mathbf{x}^{(t)}|\mathbf{x}^{\prime})}{q(\mathbf{x}^{\prime}|\mathbf{x}^{(t)})}\right\},
    \label{eq:acc_prob}
  \end{align}
} 
where $q(\mathbf{x}^{(t)}|\mathbf{x}^{\prime})$ is the reverse proposal calculated similarly as the forward proposal $q(\mathbf{x}^{\prime}|\mathbf{x}^{(t)})$. This equation is derived by substituting the target distribution $\pi$ \eqref{eq:post} and the discrete proposal $q$ \eqref{eq:dis_prop_fac1} into \eqref{eq:acc_prob_general}. Note that the indicator $\mathbb{I}$ is omitted for simplicity.

\renewcommand{\algorithmicrequire}{\textbf{Input:}}
\renewcommand{\algorithmicensure}{\textbf{Output:}}
\renewcommand{\algorithmiccomment}[1]{/* #1 */}
\newcommand{\IfThen}[2]{
  \STATE \algorithmicif\ #1\ \algorithmicthen\ #2}
\newcommand{\parfor}[1]{\STATE \algorithmicfor\ #1  \textbf{do in parallel}}
\algsetup{indent=0.5em}
\CheckRmv{
\begin{algorithm}[t]
	\caption{DMALA Sampler}
	\label{alg:dmala}
	\begin{algorithmic}[1] 
		\REQUIRE ${\bf{y}}$, ${\bf{H}}$, $\sigma^2$, $\alpha$.
		\STATE \textbf{Initialize:} $\mathbf{x}^{(1)}$, $f(\mathbf{x}^{(1)})$, $\nabla f(\mathbf{x}^{(1)})$.
      \STATE Construct the proposal $q_n(\cdot | x_{n}^{(1)}), n=1,\ldots,N$ in \eqref{eq:dis_prop_fac2}.
			
		\FOR{$t=1$ to $T-1$}
      \STATE Sample $x^{\prime}_n \sim q_n(\cdot | x^{(t)}_{n})$ for $n=1$ to $N$ in parallel. 
      \STATE Compute $q(\mathbf{x}^{\prime} | \mathbf{x}^{(t)}) = \prod_{n=1}^{N} q_n(x^{\prime}_n | x^{(t)}_{n})$.
      \STATE Compute $\nabla f(\mathbf{x}^{\prime})$ via \eqref{eq:grad}.
      \STATE Construct the proposal $q_n(\cdot| x^{\prime}_n), n=1,\ldots,N$  in \eqref{eq:dis_prop_fac2} and compute $q(\mathbf{x}^{(t)}|\mathbf{x}^{\prime} ) = \prod_{n=1}^{N} q_n(x^{(t)}_{n}| x^{\prime}_n)$.
      \STATE Compute $f(\mathbf{x}^{\prime})$ via \eqref{eq:f}.
      \STATE Compute $ A(\mathbf{x}^{\prime}|\mathbf{x}^{(t)})$ via \eqref{eq:acc_prob} and sample $u\sim \mathcal{U}(0, 1)$.
      \IF{$ A(\mathbf{x}^{\prime}|\mathbf{x}^{(t)}) > u$}
      \STATE $\mathbf{x}^{(t+1)} = \mathbf{x}^{\prime}$, $f(\mathbf{x}^{(t+1)}) = f(\mathbf{x}^{\prime})$, and $q_n(\cdot | x^{(t+1)}_{n}) = q_n(\cdot| x^{\prime}_n), n=1,\ldots,N$.
      \ELSE
      \STATE ${\mathbf{x}^{(t+1)} = \mathbf{x}^{(t)}}$, ${f(\mathbf{x}^{(t+1)}) = f(\mathbf{x}^{(t)})}$, and ${q_n(\cdot | x^{(t+1)}_{n})} = q_n(\cdot| x^{(t)}_{n}), n=1,\ldots,N$.
      \ENDIF
		\ENDFOR
		\ENSURE Samples $\{\mathbf{x}^{(t)}\}_{t=1}^{T}$.
	\end{algorithmic} 
\end{algorithm}
}

The sampler using DMALA is outlined in \algref{alg:dmala}.
Parallel samplers can be employed for sample generation, facilitating efficient utilization of computational resources while reducing correlation among the generated samples.
Utilizing these samples, hard decisions are determined by selecting the sample with the smallest residual norm $\|\mathbf{y}-\mathbf{Hx}\|$, whereas soft decisions are inferred using the Monte Carlo methods discussed in \secref{sec:basics}. In this paper, we focus on the latter approach, with the developed method detailed in \secref{sec:llr}. 

\CheckRmv{
    \begin{figure*}[t]
      \centering
      \subfigure[Existing methods]{
        \includegraphics[width=2.04in]{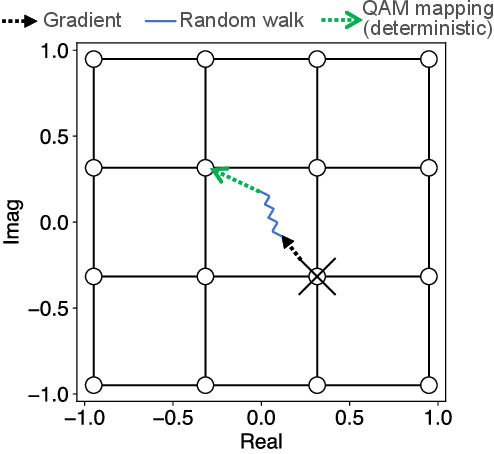}
        \label{fig:mhgd_visual}
      }
      \subfigure[DMALA]{
        \includegraphics[width=4.75in]{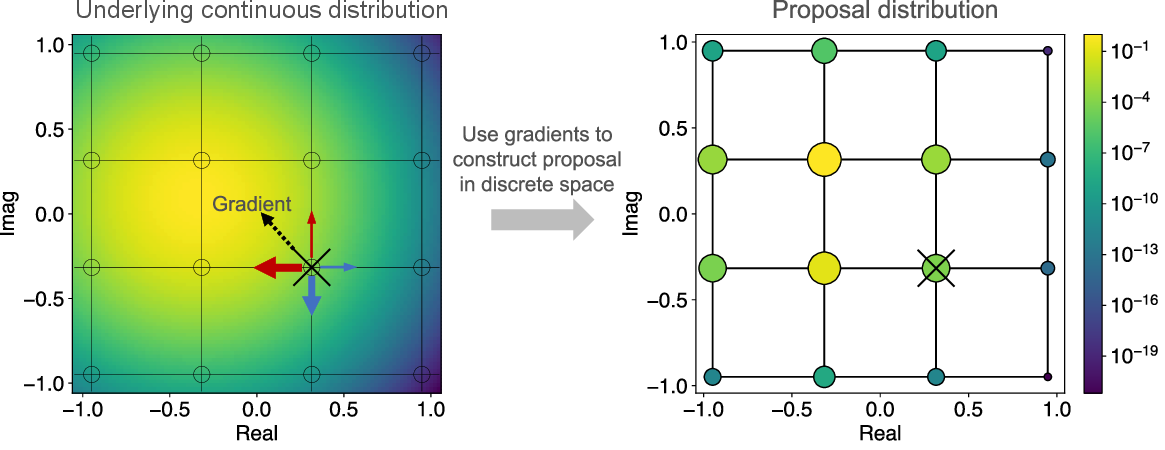}
        \label{fig:dmala_visual}
      }
      \caption{Visual comparison between existing gradient-based MCMC methods \cite{gowdaMetropolisHastingsRandomWalk2021,zhou2023gradient} and the proposed DMALA for the 16-QAM lattice. The black cross represents the current state of the Markov chain. In (a), the candidate sample is generated via a sequential process of gradient descent, random walk, and QAM mapping. In (b), the left subplot shows the underlying continuous distribution of the target discrete distribution, with colors indicating probability levels. 
      Gradients from the logarithm of this continuous distribution are employed to construct the proposal distribution, as shown by color dots in the right subplot.}
      \label{fig:visual}
    \end{figure*}
  }

\begin{remark}
  The target distribution may demonstrate strong second-order correlations across dimensions.
  In such cases, using naive gradients in \eqref{eq:grad} leads to slow convergence \cite{gowdaMetropolisHastingsRandomWalk2021,zhangLangevinlikeSamplerDiscrete2022,rhodesEnhancedGradientbasedMCMC2022}.
  {To accelerate convergence, we explore a preconditioned variant of DMALA.}
  Specifically, we precondition the naive gradients using the inverse of a damped Hessian, given by
  \CheckRmv{
    \begin{equation}
      \mathbf{M}  = (\mathbf{H}^{T}\mathbf{H} + \gamma \mathbf{I}_N)^{-1},
    \end{equation}
  }
  where $\gamma> 0$ is a Tikhonov damping parameter.
  This preconditioner $\mathbf{M}$ utilizes the second-order information from the target distribution to expedite the gradient descent process.
  With this preconditioner, the Langevin algorithm (in the continuous space) proceeds as
  \CheckRmv{
    \begin{equation}
      \mathbf{x}^{\prime} = \mathbf{x}^{(t)} + \frac{\alpha}{2} \mathbf{M}\nabla f(\mathbf{x}^{(t)}) + \sqrt{\alpha\beta} \mathbf{w}^{(t)}. 
    \end{equation}
  }
In this variant, we also scale the random perturbation with an additional parameter $\beta>0$, which allows for adjusting the gradient to perturbation intensity ratio.  Correspondingly, in the preconditioned DMALA, the proposal in \eqref{eq:dis_prop_fac2} is adapted by substituting $\mu(x^{\prime}_n)$ with $\nu(x^{\prime}_n)$, represented by  
  \CheckRmv{
    \begin{equation}
      \nu(x^{\prime}_n) = \frac{1}{2\beta}\left[\mathbf{M}\nabla f(\mathbf{x}^{(t)})\right]_n(x^{\prime}_n - x^{(t)}_n) - \frac{(x^{\prime}_n - x^{(t)}_n)^2}{2 \alpha\beta}.
    \end{equation}
  }
\end{remark}

\begin{remark}
  We present a visual comparison between existing gradient-based MCMC methods for MIMO detection \cite{gowdaMetropolisHastingsRandomWalk2021,zhou2023gradient} and the proposed DMALA, as illustrated in \figref{fig:visual}. 
  \figref{fig:mhgd_visual} depicts how existing methods relax the problem to the continuous space, where they generate a candidate sample through a sequence of steps: gradient descent, random walk (by adding a random Gaussian perturbation), and QAM mapping. Regrettably, the QAM mapping, which discretizes the continuous update to its nearest lattice point, is \textit{deterministic} and lacks an associated proposal probability. This absence makes the exact MH correction unattainable, leading, as shown later, to \textit{inexact} sampling and significant inference errors. Conversely, as depicted in \figref{fig:dmala_visual}, DMALA constructs the proposal with guidance from gradients of the underlying continuous function, allowing for an exact MH correction. This key differentiation ensures the convergence of our method, a point elaborated upon in the following subsection. 
\end{remark}

\subsection{Convergence of the DMALA Sampler}
A distinct feature of the DMALA sampler, compared to existing gradient-based MCMC methods \cite{gowdaMetropolisHastingsRandomWalk2021,zhou2023gradient}, is its guaranteed (asymptotic) convergence to the target distribution. To begin with, we analyze the statistical properties of the Markov chain induced by DMALA. 
For two states $\mathbf{x},\mathbf{x}^{\prime} \in \mathcal{A}^{N\times1}$, the transition probability $P(\mathbf{x}^{\prime}|\mathbf{x})$ of this chain is given by 
  \CheckRmv{
    \begin{align}
      &P(\mathbf{x}^{\prime}|\mathbf{x}) \nonumber\\ 
      &=\left\{\begin{array}{ll}
        q(\mathbf{x}^{\prime} | \mathbf{x}) A(\mathbf{x}^{\prime} | \mathbf{x}), & \text { if } \mathbf{x}^{\prime} \neq \mathbf{x}, \\
        q(\mathbf{x} | \mathbf{x})+\sum_{\mathbf{z} \neq \mathbf{x}} q(\mathbf{z} | \mathbf{x})\big(1-A(\mathbf{z} | \mathbf{x})\big), & \text { otherwise},
        \end{array}\right.
      \label{eq:trans_prob_1}
    \end{align}
  }
where $q(\cdot | \cdot)$ represents the proposal probability, computed as per \eqref{eq:dis_prop_fac1}, and $A(\cdot | \cdot)$ denotes the acceptance probability according to \eqref{eq:acc_prob}.
Armed with this transition probability (kernel) in position, we present a lemma detailing the statistical properties of the Markov chain, proved in Appendix \ref{appendix1}.

\begin{lemma} \label{th:property}
  Assume that the probability density function of the noise $\mathbf{n}$ satisfies $0< p_{\bf n}(\cdot)<+\infty$, i.e., $\sigma^2 > 0$. 
  Then, the transition kernel of the Markov chain induced by DMALA ($\alpha>0$) is \textit{irreducible} and \textit{aperiodic}.
  Moreover, the target posterior distribution $\pi$ is a \textit{unique stationary distribution} of this chain. 
\end{lemma}

Building upon Lemma \ref{th:property} and the convergence theorem of MCMC, we assert the theorem concerning the exponential convergence of the Markov chain induced by DMALA, as presented herein without an accompanying proof. A comprehensive proof is available in \cite[Theorem 4.9]{levinMarkovChainsMixing}.

\begin{theorem} \label{th:convergence}
  Given that Lemma \ref{th:property} holds, the Markov chain induced by DMALA exhibits exponential convergence to its unique stationary distribution $\pi$ in total variation distance. Specifically, there exist constants $C>0$ and $0<r<1$ such that
  \CheckRmv{
    \begin{equation}
      \|P^{(t)}(\mathbf{x}, \cdot) - \pi\|_{\rm TV} \leq C r^t,\; \forall \mathbf{x}\in \mathcal{A}^{N\times 1}.
    \end{equation}
  }
  Herein, $r$ represents the convergence rate \cite{chenConvergenceAnalysesComparisons2002}, with a smaller $r$ indicating faster convergence, $P^{(t)}(\mathbf{x}, \cdot)$ denotes the probability distribution induced by the $t$-step transition function of the Markov chain with the initial state $\mathbf{x}$, defined as \cite{chenConvergenceAnalysesComparisons2002}
  \CheckRmv{
    \begin{equation}
      P^{(t)}(\mathbf{x}, \mathbf{x}^{\prime}) = P\big(\mathbf{x}^{(t+1)} =\mathbf{x}^{\prime}|\mathbf{x}^{(1)} = \mathbf{x}\big),
    \end{equation}
  }
  and $\|\pi_1-\pi_2\|_{\rm TV}$ signifies the total variation distance between two distributions ${\pi}_1$ and $\pi_2$ over $\mathcal{A}^{N\times 1}$, defined as
  \CheckRmv{
    \begin{equation}
      \|\pi_1 - \pi_2\|_{\rm TV} = \frac{1}{2}\sum_{\mathbf{x}\in\mathcal{A}^{N\times1}} |\pi_1(\mathbf{x}) - \pi_2(\mathbf{x})|.
    \end{equation}
  }
\end{theorem}

Theorem \ref{th:convergence} implies that the distribution of the samples generated by DMALA exponentially converges to the target distribution for large $t$. Furthermore, this theorem sheds light on the rate of convergence, denoted as $r$. Delving into the convergence rate $r$, we consider the discrete space $\mathcal{A}^{N\times 1}$ represented by  $\left\{\mathbf{x}_{1}, \mathbf{x}_{2}, \ldots, \mathbf{x}_{Q^N}\right\}$. The transition kernel of the Markov chain, induced by DMALA, can be modeled as a $Q^N \times Q^N$ matrix $\mathbf{P} = [P_{ij}]$. Here, the entry $P_{ij}$ signifies the transition probability from $\mathbf{x}_i$ to $\mathbf{x}_j$, calculated as per \eqref{eq:trans_prob_1}. It is established that the convergence rate $r$ corresponds to the second largest eigenvalue ${\lambda_2}$ of $\mathbf{P}$ \cite{chenConvergenceAnalysesComparisons2002,kontoyiannis2012geometric}. Although deriving analytical expressions for this eigenvalue poses challenges, it is feasible to exactly compute $r$ in small sample spaces, aiding in convergence diagnostics. Such an analysis not only enhances comprehension of the algorithm's performance but also illuminates paths for further improvement.

\CheckRmv{
  \begin{figure}[t]
    \centering
    \includegraphics[width=3.1in]{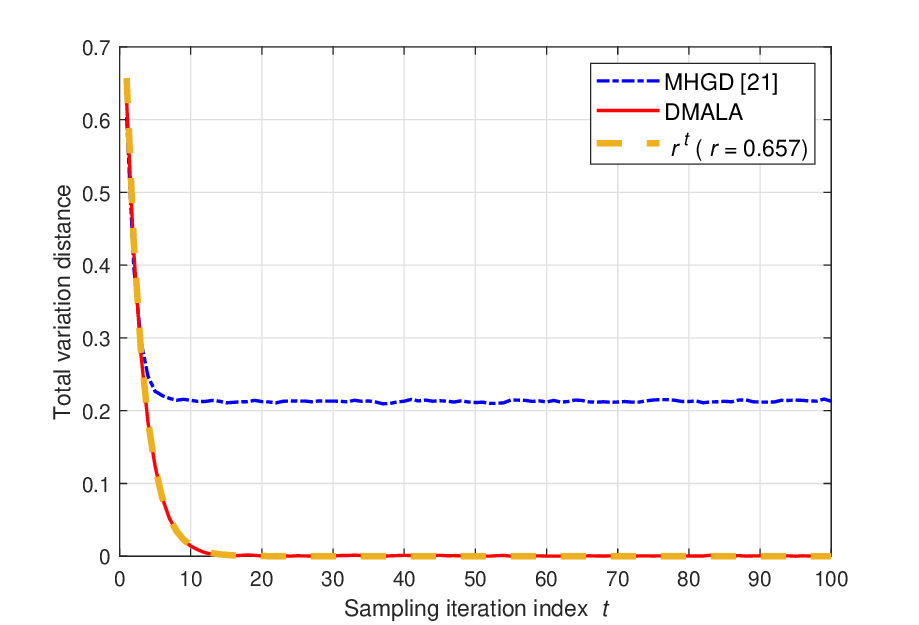}
    \caption{Total variation distance as a function of the number of sampling iterations within a \Times{2}{2} MIMO system featuring Rayleigh fading channels, QPSK modulation, and \SNR{=}{8}.}
    \label{fig:tv_theory}
  \end{figure}
}

\CheckRmv{
  \begin{figure}[t]
    \centering
    \includegraphics[width=3.1in]{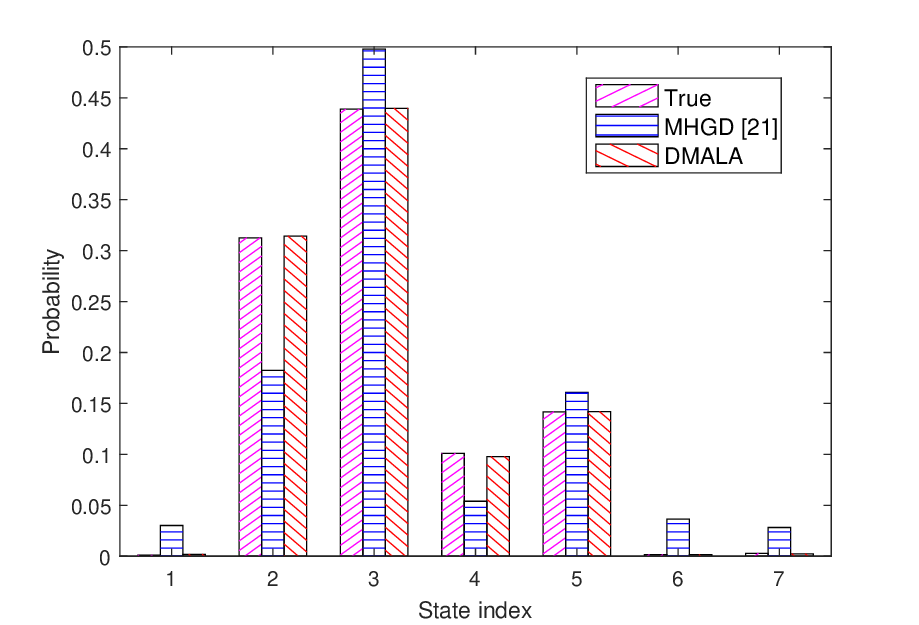}
    \caption{Comparison between the sampled and true posterior distributions within the \Times{2}{2} MIMO-QPSK space, which encompasses 16 possible states. States exhibiting a probability lower than $10^{-3}$ are omitted for clarity.}
    \label{fig:distribution}
  \end{figure}
}

We conducted an empirical verification of Theorem \ref{th:convergence} using a \Times{2}{2} MIMO system with quadrature phase shift keying (QPSK) modulation (${Q^N=2^4=16}$) and the signal-to-noise ratio (SNR) per receiving antenna of 8 dB.\footnote{{The SNR is defined as $\text{SNR}=\mathbb{E}[\|\mathbf{Hx}\|^2] / \mathbb{E}[\|\mathbf{n}\|^2]$.}} \figref{fig:tv_theory} shows the total variation distance between the sampled distribution $P^{(t)}(\mathbf{x}, \cdot)$ from DMALA and the true target distribution as a function of the sampling iteration index $t$. This sampled distribution was derived by collecting the generated samples of $10^5$ independent samplers at each time step $t$ and counting the number of occurrences to reflect the probability of each state in the \Times{2}{2} MIMO-QPSK space. For this empirical study, we employed the preconditioned version of DMALA. We benchmarked against a state-of-the-art gradient-based MCMC method for MIMO detection, specifically MHGD \cite{gowdaMetropolisHastingsRandomWalk2021}. An exponential convergence curve of $r^t$ serves as a reference.
For this specific channel realization, $r=0.657$ was determined through the eigendecomposition of the $16\times 16$ transition matrix $\mathbf{P}$ associated with DMALA. 

The analysis yields several key insights. First, the total variation distance for DMALA approaches zero, indicating successful convergence to the target distribution, whereas for MHGD, it converges to an approximate value of 0.2, suggesting a deviation from the target. This outcome underscores the superiority of DMALA's methodical probabilistic sampling approach. Second, the empirical convergence curve of DMALA's total variation distance (depicted by a solid line) closely aligns with the theoretical exponential convergence curve (dashed line), thereby corroborating the assertions of Theorem \ref{th:convergence}. Additionally, when comparing the sampled distributions to the true distribution, as shown in \figref{fig:distribution} via histograms, it is evident that MHGD's sampled distribution exhibits bias, while DMALA's distribution closely matches the true distribution.

\CheckRmv{
  \begin{figure}[t]
    \centering
    \subfigure[Naive]{
      \includegraphics[width=3in]{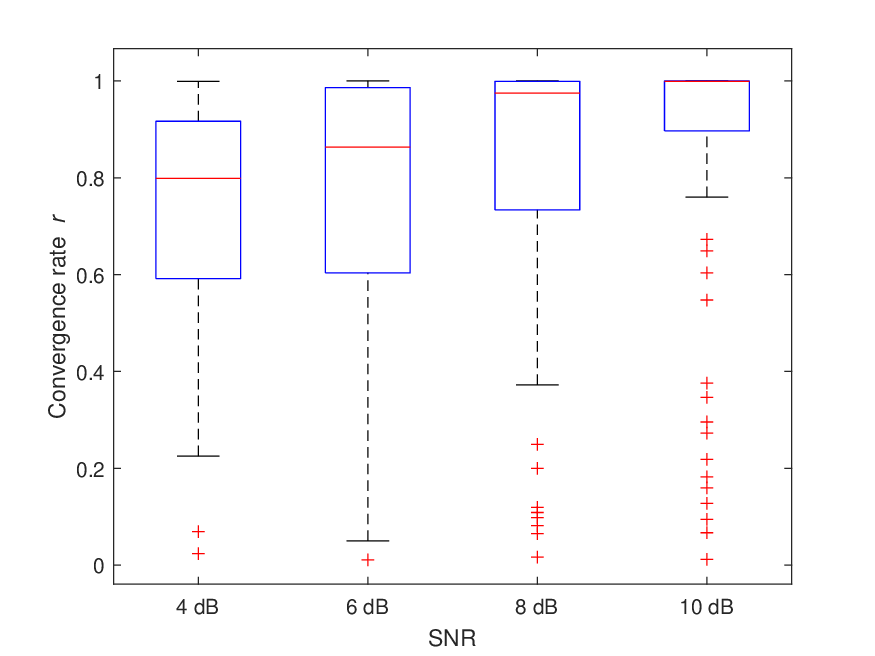}
      \label{fig:naive}
    }
    \subfigure[Preconditioned]{
      \includegraphics[width=3in]{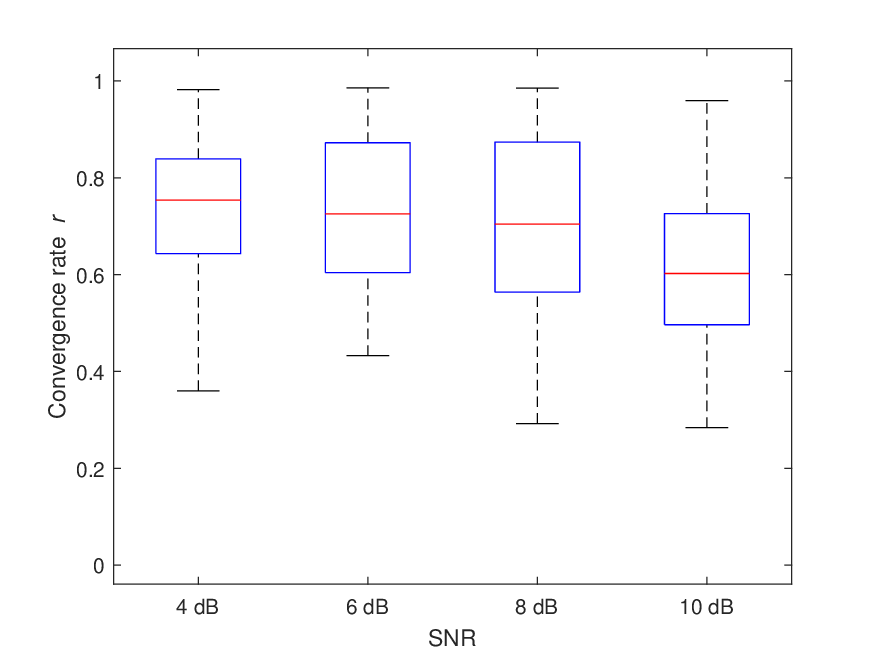}
      \label{fig:preconditioned}
    }
    \caption{Boxplots illustrating the convergence rates in a \Times{2}{2} MIMO system employing QPSK modulation under Rayleigh fading channels. The convergence rates for naive and preconditioned DMALA configurations are depicted in panels (a) and (b), respectively.}
    \label{fig:rate}
  \end{figure}
}

In \figref{fig:rate}, we extend our investigation to examine the performance differences between naive and preconditioned DMALA configurations under varying SNR conditions. The system setup remains identical to that described in \figref{fig:tv_theory}, with SNR levels adjusted to 4, 6, 8, and 10 dB. For each SNR setting, we simulate 100 independent channel realizations. 
The transition matrix $\mathbf{P}$ is constructed for each realization, from which we compute the second largest eigenvalue, i.e., the convergence rate $r$.
\figref{fig:rate} shows the boxplots\footnote{A boxplot is a method for displaying the distribution of a dataset based on the five-number summary: the minimum, first  quartile, medium, third quartile, and maximum. The first and third quartiles are denoted by the two ends of the box, and the medium is represented by the middle bar. The points outside the upper and lower bounds are considered outliners.} of these convergence rates, distinguishing between the results for naive DMALA in \figref{fig:naive} and the preconditioned variant in \figref{fig:preconditioned}.
From \figref{fig:rate}, naive DMALA exhibits increasingly slow convergence ($r \to 1$) at SNR values above 8 dB, a phenomenon recognized in the literature as the high SNR stalling issue \cite{farhang-boroujenyMarkovChainMonte2006,hedstromAchievingMAPPerformance2017}.  
Conversely, the preconditioned DMALA demonstrates steady convergence rates across the explored SNR spectrum, indicating notable enhancements in performance. 

{We further investigate the convergence rate beyond empirical analysis, drawing inspiration from \cite{wang2024randomized}. We first introduce a lemma on the proposal distribution of DMALA, with the proof provided in Appendix~\ref{appendix_lemma2}. 

\begin{lemma} \label{lemma:ratio}
  For the Markov chain of the proposed DMALA, there exists a constant $\xi>0$ such that 
  \CheckRmv{
    \begin{equation}
      \frac{q(\mathbf{x}^{\prime} | \mathbf{x}^{(t)})}{\pi(\mathbf{x}^{\prime})} \geq \xi \cdot G(\mathbf{x}^{(t)}, \mathbf{x}^{\prime}),\;\forall \mathbf{x}^{(t)}\in \mathcal{A}^{N\times 1},
    \end{equation}
  }
  where 
  \CheckRmv{
    \begin{equation}
      \xi = \frac{\sum_{\mathbf{s}\in \mathcal{A}^{N\times 1}} \exp\big( f(\mathbf{s})\big) }{\prod_{n=1}^{N} \sum_{x_n^{\prime} \in \mathbb{Z}} \exp \big(-\frac{1}{2\alpha} |x_n^{\prime}|^2 \big)}
    \end{equation} 
  }
  with $\mathbb{Z}$ denoting the set of all integers (encompassing the QAM lattice $\mathcal{A}$), and the function $G$ is given by 
  \CheckRmv{
    \begin{equation}
      G(\mathbf{x}^{(t)}, \mathbf{x}^{\prime}) = \frac{\exp \big(-\frac{1}{2\alpha} \|\mathbf{x}^{\prime} - \mathbf{x}^{(t)} - \frac{\alpha}{2}\nabla f(\mathbf{x}^{(t)}) \|^2 \big)}{\exp \big(f(\mathbf{x}^{\prime}) \big)}.
    \end{equation}
  }
\end{lemma}

Based on this lemma, we have 
\CheckRmv{
  \begin{equation}
    q(\mathbf{x}^{\prime} | \mathbf{x}^{(t)}) \geq \xi\cdot G(\mathbf{x}^{(t)},\mathbf{x}^{\prime}) \pi(\mathbf{x}^{\prime}) \geq \delta \pi(\mathbf{x}^{\prime})
    \label{eq:relationship}
  \end{equation}
}
for all $\mathbf{x}^{(t)}, \mathbf{x}^{\prime} \in \mathcal{A}^{N\times 1}$, where 
\CheckRmv{
  \begin{equation}
    \delta = \xi \cdot \underset{\mathbf{x}^{(t)}, \mathbf{x}^{\prime} \in \mathcal{A}^{N\times 1}}{\min} [G(\mathbf{x}^{(t)}, \mathbf{x}^{\prime})] > 0. 
  \end{equation}
}
Following this relationship, the coupling technique can be employed to establish the following theorem that delineates the exponential convergence rate of DMALA. The related proof is available in \cite[Theorem 1]{wang2018on}.

\begin{theorem}
  The convergence rate $r$ in Theorem~\ref{th:property} can be specified as $r=1-\delta$, where $\delta = \xi \cdot \underset{\mathbf{x}^{(t)}, \mathbf{x}^{\prime} \in \mathcal{A}^{N\times 1}}{\min} [G(\mathbf{x}^{(t)}, \mathbf{x}^{\prime})] $ also provides a lower bound for the spectral gap (i.e., $1-\lambda_2$) of DMALA's transition kernel $\mathbf{P}$.
\end{theorem}

This theorem shows that the convergence rate depends on factors such as the step size $\alpha$ and the system setup. 
Optimizing these parameters, such as selecting a suitable $\alpha$ to make $\xi$ and $G(\mathbf{x}^{(t)}, \mathbf{x}^{\prime})$ approach 1 and improve the convergence rate, is beyond the scope of this work and is left for future research.}

\subsection{LLR Computation Methods}   \label{sec:llr} 
After deriving the sample list $\mathcal{L}$, various methods can be utilized for LLR computation.
Conventional list-based soft-output detectors \cite{hochwaldAchievingNearcapacityMultipleantenna2003,guoAlgorithmImplementationKbest2006,wangExpectationPropagationBasedSampling2021} approximate the APPs in \eqref{eq:exact_llr} by using $\mathcal{L}$ to replace the entire set $\mathcal{A}^{N\times 1}$ of all possible vectors. 
Since the list size is generally much smaller than $Q^N$, the exponentially increased complexity can be alleviated. 
Consequently, the LLR as defined in \eqref{eq:exact_llr} is approximated by
\CheckRmv{
  \begin{equation}
    \hat{L}_{k}=\log \frac{\sum_{\mathbf{x}\in {\mathcal{L}} \cap \mathcal{A}^{N\times 1}_{k+}} \exp\big(f(\mathbf{x})\big)}{\sum_{\mathbf{x}\in {\mathcal{L}}\cap \mathcal{A}^{N\times 1}_{k-}} \exp\big(f(\mathbf{x})\big)}.  
    \label{eq:llr_approx_app}
  \end{equation}
}
However, this approach is not tailored to the proposed sampling-based detector since the distribution information of the generated samples is not fully harnessed. 

To better accommodate the proposed sampling algorithm, we adopt an alternative approach for LLR computation based on the IS-based Monte Carlo method \cite{farhang-boroujenyMarkovChainMonte2006,senstRaoBlackwellizedMarkovChain2011}.
Initially, to facilitate a Monte Carlo summation for approximation, we reformulate the APPs $p(b_k = \pm 1 | \mathbf{y})$  in \eqref{eq:exact_llr} into the form of expectation.
Observe that the APPs can be expressed as
\CheckRmv{
  \begin{align}
    p(b_{k}=\pm 1 | \mathbf{y})
    &=\sum_{\mathbf{b}_{-k}} p(b_{k}=\pm 1,\mathbf{b}_{-k} | \mathbf{y}) \nonumber\\
    &=\sum_{\mathbf{b}_{-k}} p(b_{k}=\pm 1 | \mathbf{y}, \mathbf{b}_{-k}) p(\mathbf{b}_{-k} |\mathbf{y}), 
  \end{align}
}
where $p(\mathbf{b}_{-k}|\mathbf{y})$ is considered as $g(x)$ and $p(b_k=\pm 1|\mathbf{y}, \mathbf{b}_{-k})$ as $h(x)$ in \eqref{eq:expectation}. Thus, the computation of APPs translates to evaluating $\mathbb{E}_{p(\mathbf{b}_{-k}|\mathbf{y})}[p(b_k=\pm 1|\mathbf{y}, \mathbf{b}_{-k})]$, where IS-based Monte Carlo summation becomes applicable.

Furthermore, to implement the IS-based method, we introduce an auxiliary distribution as the target according to \cite{senstRaoBlackwellizedMarkovChain2011}:
\CheckRmv{ 
  \begin{equation}
    \pi_{\rm a}(\mathbf{x}) \propto \exp\big({f(\mathbf{x})/\tau}\big)\prod_{n=1}^N \mathbb{I}_{x_n \in \mathcal{A}},
    \label{eq:aux}
  \end{equation}
}
where $\tau>1$ is a temperature parameter for enhancing the mobility of the Markov chain \cite{farhang-boroujenyMarkovChainMonte2006,wangExpectationPropagationBasedSampling2021,hassibiOptimizedMarkovChain2014}. 
Given that $\pi_{\rm a}$ represents a tempered posterior and maintains a form analogous to the posterior distribution $\pi$ in \eqref{eq:post}, our proposed DMALA is well-suited to sample from it, following the derivations in \secref{sec:algorithm}. 

Let $\mathcal{L} = \{\mathbf{x}^{[s]}\}_{s=1}^S$ denote the sample list utilized for LLR computation, where $S$ is the size of the sample list (including repetitive samples). 
To reflect our focus on employing samples nearing convergence for decisions rather than all samples generated during the sampling iterations, we shift the index notation from $(t)$ to $[s]$.
Building upon \eqref{eq:is_mc} and \eqref{eq:aux} and through some algebraic manipulation, the computation of \eqref{eq:exact_llr} can be approximated by
\CheckRmv{
  \begin{equation}
    \hat{L}_k=\log\frac{\sum_{s=1}^S\frac1{1+\exp(-\gamma_{k}^{[s]})}\exp\left(\frac{\tau-1}\tau f(\mathbf{x}_{+1}^{[s]})\right)}{\sum_{s=1}^S\frac1{1+\exp(+\gamma_{k}^{[s]})}\exp\left(\frac{\tau-1}\tau f(\mathbf{x}_{-1}^{[s]})\right)},
    \label{eq:llr_is}
  \end{equation}
}
where $\gamma_{k}^{[s]} = \frac{1}{\tau}\left(f(\mathbf{x}_{+1}^{[s]}) - f(\mathbf{x}_{-1}^{[s]})\right)$, and $\mathbf{x}_{+ 1}^{[s]}$ and $\mathbf{x}_{- 1}^{[s]}$ are mapped from $\mathbf{b}^{[s]}$ with its $k$-th bit ${b}_k^{[s]}$ corresponding to $+1$ or $-1$, respectively. The bit vector $\mathbf{b}^{[s]}$ is  demapped from the sample $\mathbf{x}^{[s]}$ drawn from $\pi_{\rm a}$ by DMALA. 
A detailed derivation of \eqref{eq:llr_is} is provided in Appendix~\ref{appendix2}.
{The complete process of DMALA-based soft MIMO detection, which integrates the proposed DMALA sampler and IS-based LLR computation, is detailed in \algref{alg:complete}.}

\CheckRmv{
{\begin{algorithm}[t]
	\caption{{DMALA-Based Soft MIMO Detection}}
	\label{alg:complete}
	{\begin{algorithmic}[1] 
		\REQUIRE ${\bf{y}}$, ${\bf{H}}$, $\sigma^2$, $\alpha$, $\tau$.
	  \STATE Run the DMALA sampler (\algref{alg:dmala} with $f(\cdot)$ replaced by $f(\cdot)/\tau$) to sample from $\pi_{\rm a}(\mathbf{x})$ in \eqref{eq:aux}.   
	  \STATE Select the converged samples $\mathcal{L} = \{\mathbf{x}^{[s]}\}_{s=1}^{S}$ for LLR computation. 
    \STATE Demap $\{\mathbf{x}^{[s]}\}_{s=1}^{S}$ to derive $\{\mathbf{b}^{[s]}\}_{s=1}^{S}$.
    \STATE Find $\{\mathbf{x}_{+1}^{[s]}\}_{s=1}^{S}$ and $\{\mathbf{x}_{-1}^{[s]}\}_{s=1}^{S}$ based on $\{\mathbf{b}^{[s]}\}_{s=1}^{S}$.
    \FOR{$k=1$ to $N\log_2 Q$}
      \STATE Compute $\gamma_{k}^{[s]} = \frac{1}{\tau}\left(f(\mathbf{x}_{+1}^{[s]}) - f(\mathbf{x}_{-1}^{[s]})\right),s=1,\ldots,S$.
      \STATE Compute $\hat{L}_k$ based on \eqref{eq:llr_is}.
    \ENDFOR
      \ENSURE LLRs $\{\hat{L}_k\}_{k=1}^{N\log_2 Q}$.
	\end{algorithmic}} 
\end{algorithm}}
}

\begin{remark} 
  In alignment with established practices within the MCMC literature, we run the samplers for a sufficiently large $T$ to ensure the Markov chains have (approximately) converged. Following this, we gather a list of converged samples (typically a limited number) for LLR computation. This strategy aids in excluding potentially inaccurate samples from the initial burn-in phase \cite{farhang-boroujenyMarkovChainMonte2006,chenConvergenceAnalysesComparisons2002} and also reduces LLR computation complexity. Moreover, we observe that utilizing independent samples from parallel samplers significantly enhances the accuracy of LLR computations over using correlated samples from a single sampler. Therefore, we opt to gather converged samples from a set of parallel samplers. Although each sampler necessitates a burn-in period, this strategy can be more efficient than extending the run time of a single sampler well beyond the necessary $T$ for convergence to reduce sample correlation.
\end{remark}

\begin{remark}  
  We also evaluated the LLR computation using Monte Carlo summation in \eqref{eq:mc} with samples drawn from $\pi$.
  We found that this method requires a larger sample size to match the accuracy afforded by the approximation in \eqref{eq:is_mc}. 
  This observation aligns with the findings in \cite{senstRaoBlackwellizedMarkovChain2011} and can be ascribed to the variance reduction effect inherent in IS. Nonetheless, both methods are theoretically guaranteed to asymptotically converge to the exact LLR calculation in \eqref{eq:exact_llr}, a consequence of the law of large numbers. Furthermore, unlike the approach in \eqref{eq:llr_approx_app}, these methods not only capitalize on the samples collected in $\mathcal{L}$ but also leverage the occurrence frequencies of these samples, effectively utilizing the probability distribution across the sample space. This aspect of the methodology contributes to performance improvements, as evidenced by our simulation results.
\end{remark}

\subsection{Computational Complexity}
We analyze the computational complexity of our proposed detector, focusing on the number of dominant arithmetic operations involved,
including multiplications and exponential functions. 
The analysis is structured into two primary segments: sampling and LLR computation.
In this analysis, we denote the number of sampling iterations as $T$, the number of parallel samplers as $N_{\rm p}$, and the number of samples selected for LLR computation as $S$. 

\subsubsection{Sampling Complexity}
The computational cost during the sampling stage is twofold: initialization and iteration per sample generation.
The initialization cost of the preconditioned DMALA is dominated by the matrix inversion required to calculate the preconditioner $\mathbf{M}$, leading to $\mathcal{O}(N^3)$ real number multiplications.
Nonetheless, this preconditioner needs to be computed only once, allowing for reuse across all sampling iterations, which mitigates its impact on overall complexity. 

Each sampling iteration's complexity primarily involves three matrix-vector multiplications: two for computing the gradient $\nabla f$ as shown in \eqref{eq:grad}, and one for applying the preconditioner to the gradient as $\mathbf{M}\nabla f$, resulting in $\mathcal{O}(N^2+MN)$ multiplications. 
Specifically, the matrix-vector multiplication $\mathbf{Hx}^{\prime}$ in the gradient computation can be reduced to a summation of $\mathbf{H}$'s columns scaled by distinct QAM magnitudes since the elements of $\mathbf{x}^{\prime}$ are from $\mathcal{A}$. Additional per-iteration operations, such as proposal construction and probability calculation, impose a modest $\mathcal{O}(NQ)$ on the computational load. 

Collectively, the total complexity in the sampling stage is $\mathcal{O}\big(N^3 + (N^2+MN+NQ)TN_{\rm p}\big)$, effectively curtailing the exponential rise in complexity with increases in $N$ and $Q$. Furthermore, the parallelization of sampling across multiple samplers significantly diminishes computation delay. 

\subsubsection{LLR Computation Complexity} 
When using the IS-based LLR computation described in \eqref{eq:llr_is}, the computational demand for each bit includes the evaluation of $\gamma_k^{[s]}$ and $4S$ exponential operations.
Specifically, the calculation of $\gamma_k^{[s]}$ involves $f(\mathbf{x}_{+1}^{[s]})$  and $f(\mathbf{x}_{-1}^{[s]})$, with at least one of these evaluations already performed in the sampling stage. The remaining evaluation can also be executed efficiently, as $\mathbf{x}_{+1}^{[s]}$ and $\mathbf{x}_{-1}^{[s]}$ differ in only one entry \cite{larawayImplementationMarkovChain2009}. 
To further reduce the need for exponential function computations, \eqref{eq:llr_is} can be reformulated as
\CheckRmv{
  \begin{align}
    \hat{L}_k=&\log\left[\sum_{s=1}^{S}\exp\left(\frac{\tau-1}{\tau}f(\mathbf{x}_{+1}^{[s]}) - F(-\gamma_k^{[s]})\right)\right] \nonumber \\
    &-\log\left[\sum_{s=1}^{S}\exp\left(\frac{\tau-1}{\tau}f(\mathbf{x}_{-1}^{[s]}) - F(+\gamma_k^{[s]})\right)\right],
    \label{eq:llr_simple}
  \end{align}
}
where ${F(a) = \log(1+e^{a}),\;a\in\mathbb{R}}$. For $a\leq 0$, $F(a)$ can be approximated using pre-computed values in a lookup table; while for ${a>0}$, the identity ${F(a) = a + F(-a)}$ is utilized \cite{senstRaoBlackwellizedMarkovChain2011}. On this basis, the computation of $v=\log \sum_{s=1}^S e^{a_s}$ in \eqref{eq:llr_simple} is simplified using the following recursive operation for $s=1$ to $S$, initializing $v$ as $-\infty$  
\cite{hochwaldAchievingNearcapacityMultipleantenna2003}:
\CheckRmv{
  \begin{equation}
    v\leftarrow \log(e^v+e^{a_s}) = \max(v,a_s)+ \underbrace{\log(1+e^{-|v-a_s|})}_{F(-|v-a_s|)}.
  \end{equation}
}

\section{Simulation Results}
In this section, we demonstrate the near-optimal bit error rate (BER) performance achieved by our proposed DMALA-based detector within coded MIMO systems. Initially, we outline the common configurations employed across our simulation studies. Subsequently, we delve into the assessment of coded BER performance, exploring a variety of MIMO dimensions, channel models, and scenarios with imperfect CSI.
For the sake of conciseness, we refer to our proposed detector simply as ``DMALA'' throughout this section.

\subsection{Simulation Setups}
In our simulations, we consistently employ a rate-$3/4$ low-density parity-check code with a block length of 1944 bits for channel coding, alongside belief propagation for channel decoding.
The BER is assessed over 70,000 blocks transmitted across independent channel realizations. 
Our primary focus is on Rayleigh fading channels, characterized by independently Gaussian-distributed channel coefficients with zero mean and unit variance.  
Furthermore, to ascertain the robustness of our proposed DMALA detector, we extend our analysis to include different channel models. This encompasses Kronecker spatially correlated channels \cite{loykaChannelCapacityMIMO2001} and the more practical {3rd generation partnership project (3GPP)} technical reports (TR) 36.873 3D channels \cite{3gpp36873r12-2017}, both of which are prevalent in the literature for evaluating MIMO detectors. 
Unless explicitly noted otherwise, we assume perfect receiver knowledge of the channel matrix $\mathbf{H}$.

In light of its demonstrated superiority in \secref{sec:dmala}, we opt for the preconditioned DMALA for our evaluation.
For its parameter setups, the step size is set to $\alpha=\sigma^2$. 
Following \cite{gowdaMetropolisHastingsRandomWalk2021}, the perturbation scaling parameter $\beta$ and the damping parameter $\gamma$ are selected as $\beta  = {d_{\min}^2}/{{\sigma^2}}$ and $\gamma =  {\sigma^2}/({2d_{\min}^2})$, respectively, where $d_{\min}$ represents half of the minimum distance between any two QAM lattice points. {Benchmark algorithms for comparative analysis encompass the expectation propagation (EP) detector \cite{cespedesProbabilisticMIMOSymbol2018}, the $K$-best SD detector \cite{guoAlgorithmImplementationKbest2006}, the state-of-the-art MHGD detector \cite{gowdaMetropolisHastingsRandomWalk2021}, and the optimal MAP detector that calculates exact LLRs using \eqref{eq:exact_llr}.}
The EP detector is configured to run for 10 iterations as per \cite{cespedesProbabilisticMIMOSymbol2018} with LLRs calculated from posterior marginals approximated by Gaussian distributions \cite[Eq.~(3)]{zhouGraphNeuralNetworkEnhanced2023}.
Given the deterministic nature of its Gaussian approximation, further increases in iterations do not enhance performance.
For the $K$-best SD, we set a large list size (denoted by $K$) to establish a near-optimal baseline and utilize \eqref{eq:llr_approx_app} for LLR computation due to its deterministic list-based nature.
{For both MHGD and DMALA, $N_{\rm p}$ parallel samplers are employed, each initialized randomly and running for $T$ iterations, with the \textit{final} sample from each sampler selected for soft decisions.} Therefore, the parameter $N_{\rm p}$ also indicates the number of samples used in LLR computation. The rationale for this selection strategy has been discussed in \secref{sec:llr}. 
IS-based LLR computation is applied to both MHGD and DMALA, with a temperature parameter $\tau=2$ as suggested by \cite{senstRaoBlackwellizedMarkovChain2011}, unless specified otherwise.

\subsection{Small- and Medium-Sized MIMO}

\CheckRmv{
  \begin{figure}[t]
    \centering
    \includegraphics[width=3in]{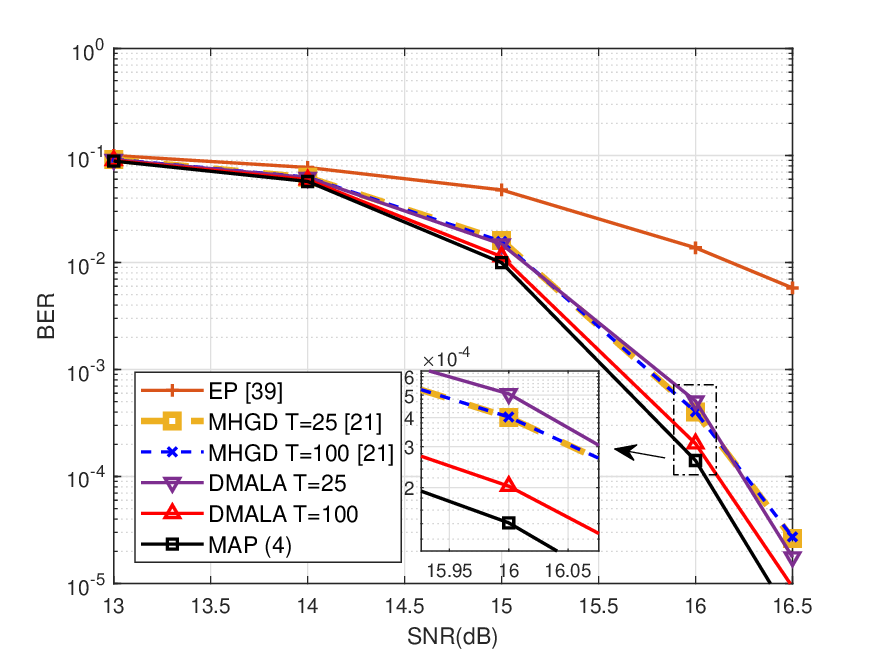}
    \caption{BER performance of DMALA across varying numbers of sampling iterations within a \Times{4}{4} MIMO system employing 16-QAM modulation under Rayleigh fading channels. Both MHGD and DMALA are configured with $N_{\rm p}=128$ parallel samplers.}
    \label{fig:4x4}
  \end{figure}
}

\figref{fig:4x4} illustrates the BER performance of DMALA with different numbers of sampling iterations $T\in\{25, 100\}$ in a \Times{4}{4} MIMO system with 16-QAM and Rayleigh fading channels. The number of parallel samplers is set as $N_{\rm p}=128$.
A notable trend observed is the gradual improvement in DMALA's BER as $T$ increases, a phenomenon that resonates with DMALA's convergence properties detailed in Theorem \ref{th:convergence}. This improvement underscores the enhanced accuracy in Monte Carlo approximation for LLRs, as the sampled distribution increasingly aligns with the target distribution with larger $T$.
Nonetheless, the gains become marginal as the performance approaches that of the optimal detector.
Contrarily, MHGD's performance does not exhibit similar improvements with increased $T$, and its best achievable BER is inferior to the optimal detector. 
This result is in accordance with the findings in \figref{fig:tv_theory}, where MHGD's sampled distribution consistently exhibits bias compared to the target distribution, hence leading to an insurmountable gap in soft decisions.

For subsequent simulations, we set $T=100$, unless noted otherwise, to ensure ample convergence of the Markov chain using DMALA, affirming the performance gains achieved via precise posterior sampling. For a fair comparison, we also set $T=100$ for MHGD.  This setup offers a balance between achieving high accuracy and maintaining moderate computational complexity, as each iteration's complexity is merely $\mathcal{O}(N^2)$.\footnote{Similar to DMALA, the per iteration complexity of MHGD is dominated by gradient computation, which is on the order of $\mathcal{O}(N^2)$.}

\CheckRmv{
  \begin{figure}[t]
    \centering
    \subfigure[QPSK]{
      \includegraphics[width=3in]{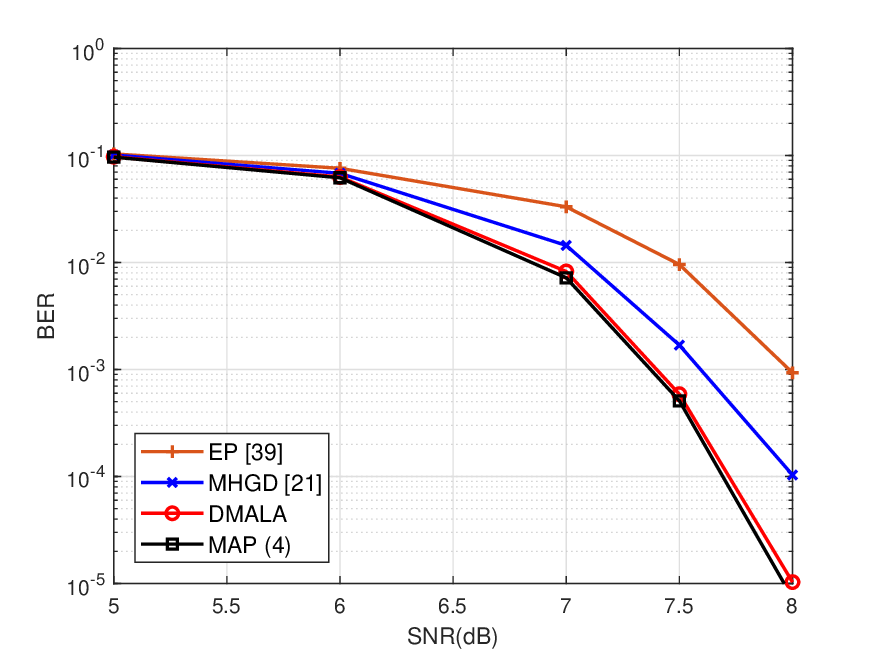}
      \label{fig:8x8_qpsk}
    }
    \subfigure[{64-QAM}]{
      \includegraphics[width=3in]{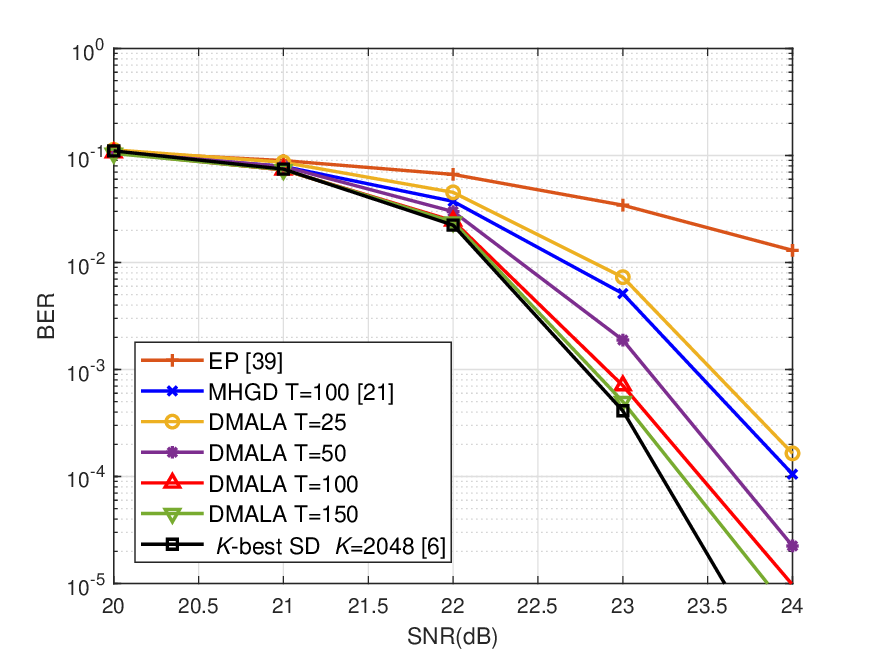}
      \label{fig:8x8_64qam}
    }
    \caption{BER performance within an \Times{8}{8} MIMO system employing QPSK/64-QAM modulation under Rayleigh fading channels. Both MHGD and DMALA are configured with $N_{\rm p}=128$ parallel samplers.}
    \label{fig:8x8}
  \end{figure}
}

\figref{fig:8x8} explores the BER performance in a medium-sized \Times{8}{8} MIMO system across different modulation schemes under Rayleigh fading channels. 
As shown in \figref{fig:8x8_qpsk}, DMALA significantly outperforms both the EP and MHGD detectors and approaches the optimal detector's performance with QPSK modulation. 
Such gains are attributed to DMALA's efficacious sampling from the target distribution, bolstering the accuracy of statistical inference. Given DMALA's comparable complexity to MHGD and substantially lower complexity relative to the optimal detector, the proposed method demonstrates substantial promise for efficient and accurate MIMO detection.

Additionally, in the context of 64-QAM modulation, as shown in \figref{fig:8x8_64qam}, the optimal MAP detector is computationally prohibitive.
{Consequently, the $K$-best SD detector, with a large list size of $K=2048$, is used as a near-optimal surrogate. We present the BER performance of DMALA with various numbers of sampling iterations $T\in \{25,50,100,150\}$ under this high-order modulation. The analysis reveals that the BER of DMALA converges quickly as $T$ increases. Notably, the DMALA detector outperforms the MHGD detector using only half the iterations ($T=50$ vs. $T=100$) and approaches the near-optimal $K$-best SD detector with $T=150$ iterations. These results demonstrate that increasing the QAM order does not significantly raise the required number of iterations for convergence, highlighting the robustness and efficiency of the proposed DMALA in handling complex modulation scenarios.}

\CheckRmv{
  \begin{figure}[t]
    \centering
    \includegraphics[width=3in]{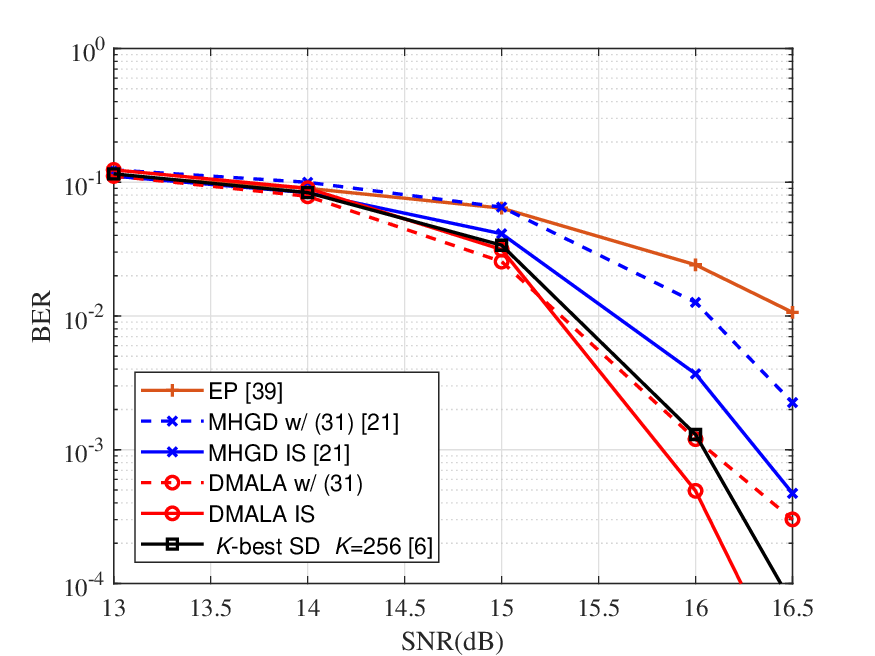}
    \caption{BER performance comparison within a \Times{16}{16} MIMO system employing 16-QAM modulation under Rayleigh fading channels. Both MHGD and DMALA are configured with $N_{\rm p}=128$ parallel samplers.}
    \label{fig:16x16}
  \end{figure}
}

\subsection{Large MIMO}
This subsection delves into the performance evaluation of the proposed detector within MIMO systems featuring an increased antenna count.
\figref{fig:16x16} showcases a performance comparison in a {large} \Times{16}{16} MIMO system using 16-QAM under Rayleigh fading channels. 
This comparison specifically focuses on the performance distinction between different LLR computation methods as outlined in \secref{sec:llr}: the conventional method, as per \eqref{eq:llr_approx_app} (``w/ \eqref{eq:llr_approx_app}''), and the IS-based method, as per \eqref{eq:llr_is}. 
These methods are evaluated for both MHGD and DMALA detectors, each utilizing $N_{\rm p}=128$ parallel samplers. 
Notably, the IS-based method outperforms the conventional method for both detectors, particularly at high SNRs. 
This advantage is largely due to the comprehensive exploitation of distribution information from the generated samples in Monte Carlo approximation, as opposed to solely relying on distinct samples within the sample list. Furthermore, DMALA's performance benefits become even more evident within this {large MIMO} setting. Specifically, the DMALA with IS significantly outperforms the $K$-best SD with $K=256$ while employing merely half the number of samples (128 vs. 256), underscoring the efficiency of sampling from critical regions of the sample space.

\CheckRmv{
  \begin{figure}[t]
    \centering
    \includegraphics[width=3in]{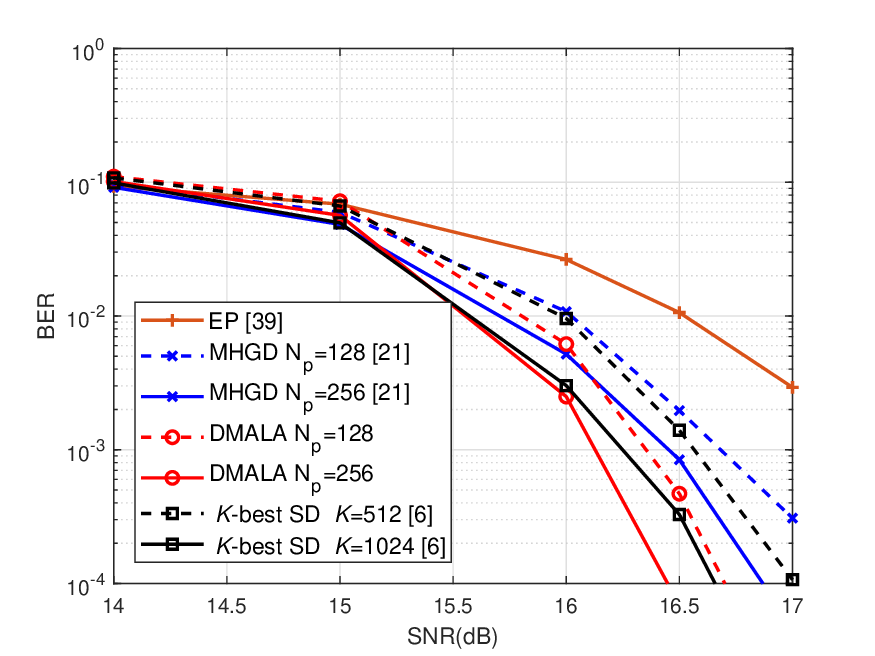}
    \caption{BER performance comparison within a \Times{24}{24} MIMO system employing 16-QAM modulation under Rayleigh fading channels.}
    \label{fig:24x24}
  \end{figure}
}

\figref{fig:24x24} shows the BER performance of DMALA in an even larger \Times{24}{24} MIMO system, again with 16-QAM. 
{We consider this system setup because enabling support for more than 24 spatial streams is a key objective in advancing MIMO technologies for the 5.5G era and has been actively pursued in numerous practical massive MIMO testbeds \cite{suyama2014evaluation,groschel2019ultra}. This configuration reflects a balanced increase in spatial streams and complexity that is representative of advanced MIMO configurations, while still being feasible within the constraints of current computational resources and hardware capabilities.} The analysis compares MHGD, DMALA, and $K$-best SD under two different parameter settings. Unsurprisingly, all methods exhibit performance improvements with an increased sample list size ($N_{\rm p}$ and $K$) for soft decision-making.  
The figure distinctly illustrates DMALA's superior performance over the benchmark methods across both parameter sets. Notably, DMALA maintains the same number of sampling iterations $T$ as previously demonstrated to achieve these gains, whereas $K$-best SD necessitates a substantial increase in $K$ to sustain competitive performance. This distinction highlights the remarkable scalability of our proposed approach.

\CheckRmv{
  \begin{figure}[t]
    \centering
    \includegraphics[width=3in]{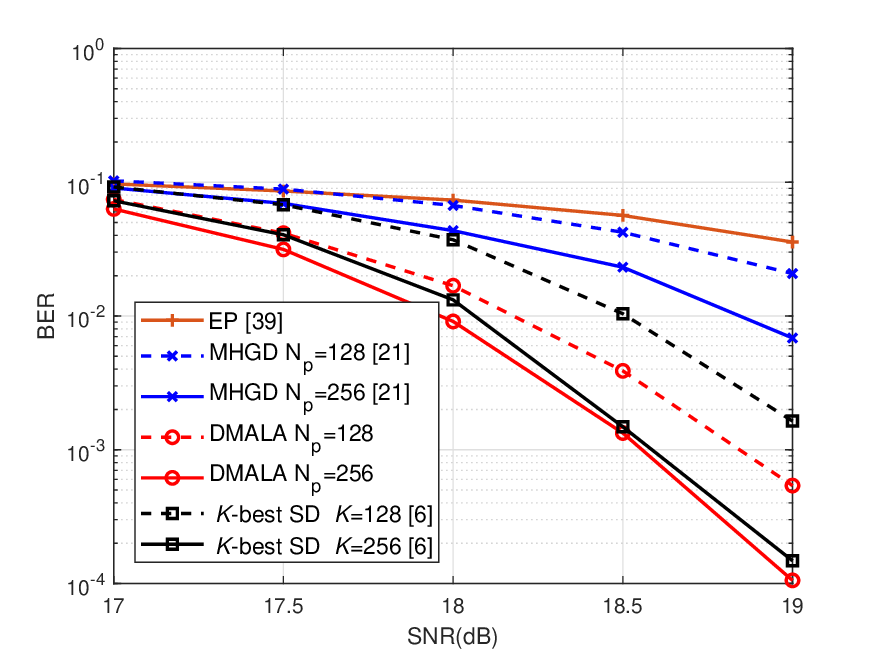}
    \caption{BER performance comparison within a \Times{16}{16} MIMO system employing 16-QAM modulation under spatially correlated channels ($\rho=0.5$).}
    \label{fig:16x16_corr}
  \end{figure}
}

\subsection{Correlated MIMO Channels}
This subsection delves into the detection performance within Kronecker spatially correlated MIMO channels, following the model introduced in \cite{loykaChannelCapacityMIMO2001}. The channel model is defined as
\CheckRmv{
  \begin{equation}
    \mathbf{H} = \mathbf{R}_{\rm r}^{1/2} \mathbf{G} \mathbf{R}_{\rm t}^{1/2},
  \end{equation}
} 
where $\mathbf{G}$ represents a Rayleigh fading channel, while $\mathbf{R}_{\rm r}$ and $\mathbf{R}_{\rm t}$ denote the spatial correlation matrices at the receiver and transmitter, respectively. The exponential correlation model \cite{loykaChannelCapacityMIMO2001} is utilized to generate these correlation matrices, with the correlation coefficient denoted by $\rho$. 

For our simulations, we consider a \Times{16}{16} MIMO system employing 16-QAM modulation, setting the spatial correlation coefficient as $\rho=0.5$.
{The selection of 0.5 as the spatial correlation coefficient represents a typical value indicating a moderate level of spatial correlation. This level is commonly encountered in practical MIMO systems and is widely adopted in the literature to assess the effectiveness of MIMO detectors \cite{he2020model,wei2020learned}.}
We explore different sample list sizes (128 and 256) for MHGD, DMALA, and $K$-best SD.  
\figref{fig:16x16_corr} shows the comparative performance analysis.
In this challenging channel scenario, all detectors undergo performance degradation---approximately a 3 dB loss in SNR compared to their performance in uncorrelated Rayleigh fading channels (as shown in \figref{fig:16x16}). Despite this, the proposed DMALA consistently achieves performance improvements over the baseline detectors with varying sample sizes, underscoring its robustness to spatial correlation effects.

\CheckRmv{
  \begin{figure}[t]
    \centering
    \includegraphics[width=3in]{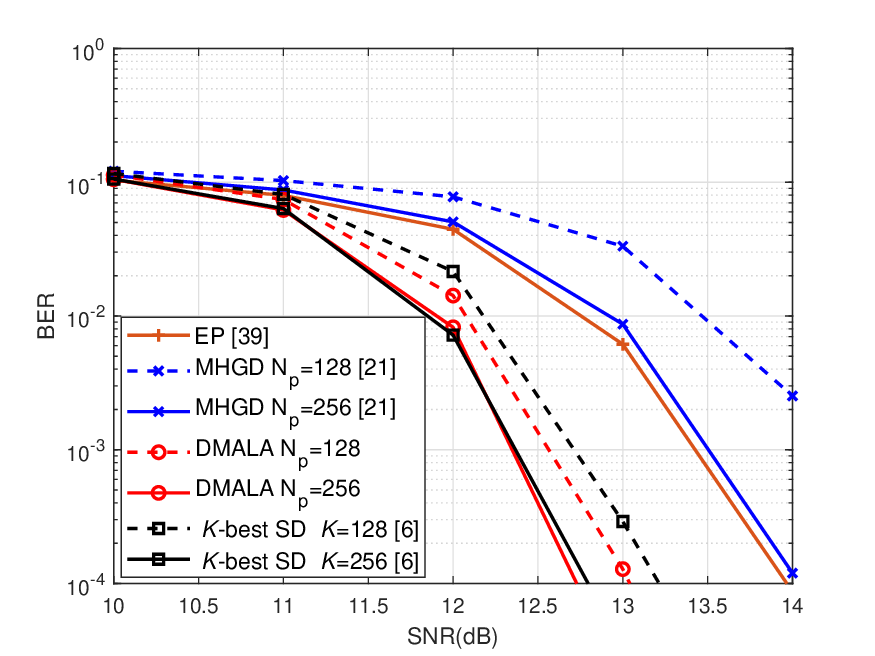}
    \caption{{BER performance comparison under the 3GPP 3D channel model employing 16-QAM modulation, 64 BS antennas, and 16 single-antenna users, giving an effective \Times{64}{16} MIMO system.}}
    \label{fig:64x16_3gpp}
  \end{figure}
}

\subsection{{3GPP Massive MIMO Channels}}
{To further attest to the effectiveness of our proposed detector, we delve into performance validation within practical 3GPP 3D MIMO channels \cite{3gpp36873r12-2017}, employing the QuaDRiGa channel simulator (Version 2.2.0) \cite{jaeckelQuaDRiGaQuasiDeterministic} for this purpose. 
We simulate the non-line-of-sight massive MIMO uplink transmission ($\nr\gg\nt$) in an urban macrocell environment, where the base station (BS) is equipped with 32 dual-polarized ($N_{\rm r} = 64$) antennas, arranged in a planar array with half-wavelength spacing and positioned at a height of 25 m.  
The BS services a cell sector extending to a 500 m radius, with 16 single-antenna users uniformly dispersed within this area, resulting in an effective $\nr\times\nt=\text{64}\times\text{16}$ MIMO system configuration.
Channel sampling is performed at a center frequency of 2.53 GHz, with transformation to the frequency domain via 256 effective subcarriers within a 20 MHz bandwidth. A total of 600 channel realizations are generated, each characterized by independent user locations, creating a channel dataset comprising \Times{600}{256} channel matrices for performance evaluation. 

\figref{fig:64x16_3gpp} presents the performance comparison within these 3GPP MIMO channels. Notably, the MHGD detector underperforms, even lagging behind the deterministic EP method, indicative of MHGD's sampled distribution deviating significantly from the target distribution in realistic channel conditions, thereby precipitating considerable errors in final inference (LLR computation). In contrast, the DMALA detector, thanks to its precise sampling capabilities, registers significant enhancements in performance relative to both MHGD and EP.  Remarkably, compared to the near-optimal $K$-best SD, DMALA achieves a 0.2 dB performance gain when using 128 samples and maintains comparable performance when using 256 samples. These findings affirm the resilience of our proposed detector in realistic massive MIMO channels.}

\subsection{Imperfect CSI}
This subsection delves into the impact of imperfect CSI on the efficacy of our proposed detector.
For the evaluation, we incorporate a linear MMSE channel estimator at the receiver to model the imperfect CSI, resulting in an estimated channel matrix 
    $\hat{\mathbf{H}} = \mathbf{H} + \mathbf{E}$, 
where $\mathbf{E}$ is the channel estimation error matrix with entries following $\mathcal{N}(0, \sigma_{\rm e}^2)$, with $\sigma_{\rm e}^2$ denoting the error variance \cite{weberImperfectChannelstateInformation2006}.
\figref{fig:ce} presents the BER performance across different levels of $\sigma_{\rm e}^2$, employing the normalized mean square error (NMSE) as a metric on the $x$-axis, defined by
\CheckRmv{
  \begin{equation}
    \text{NMSE} = \frac{\mathbb{E}[\|\mathbf{E}\|_F^2]}{\mathbb{E}[\| \mathbf{H}\|_F^2]}.
  \end{equation}
} 
Herein, we consider a \Times{16}{16} MIMO system under Rayleigh fading channels with 16-QAM modulation at \SNR{=}{16.5}. 
Consistent with the results observed under perfect CSI, the DMALA-based detector notably outperforms baseline detectors, including the $K$-best SD configured with $K=256$, across various levels of CSI accuracy.
Specifically, DMALA achieves equivalent BER performance ($\text{BER}=10^{-3}$) as MHGD and $K$-best SD at 4 dB and 1 dB higher channel estimation NMSE, respectively.
This comparative analysis unequivocally showcases the robustness of our proposed detector to imperfect CSI.

\CheckRmv{
  \begin{figure}[t]
    \centering
    \includegraphics[width=3in]{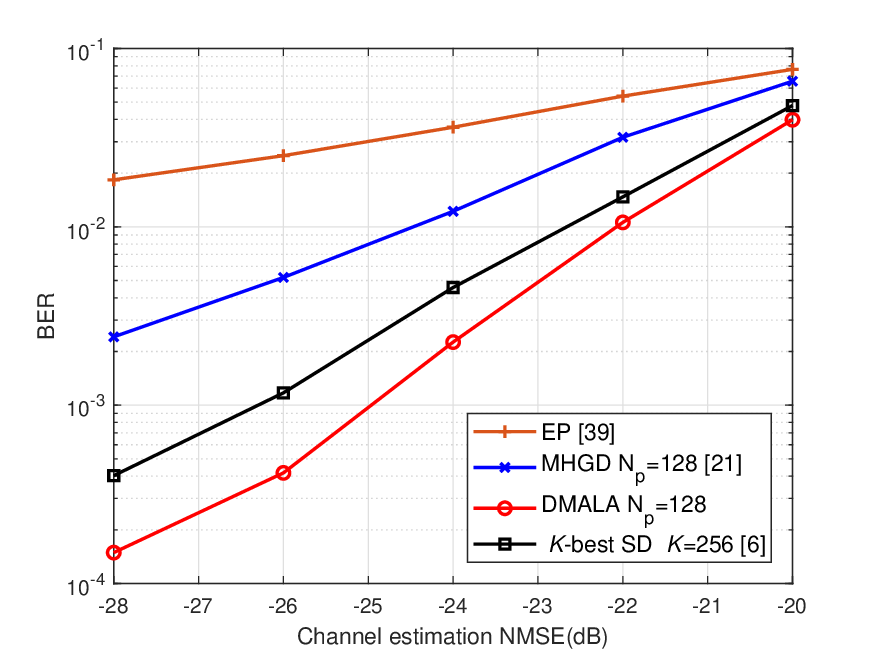}
    \caption{BER performance with respect to channel estimation errors within a \Times{16}{16} MIMO system under Rayleigh fading channels with 16-QAM modulation at \SNR{=}{16.5}.}
    \label{fig:ce}
  \end{figure}
}

{\subsection{Performance and Complexity Comparisons with Various Baselines}

\CheckRmv{
  \begin{figure}[t]
    \centering
    \subfigure[{\Times{4}{4} MIMO}]{
      \includegraphics[width=3in]{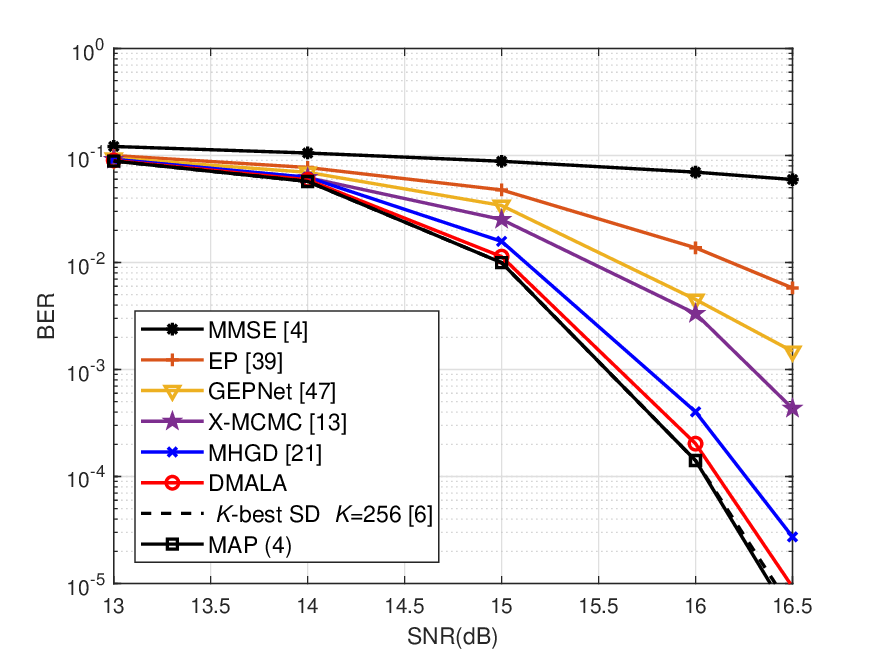}
      \label{fig:4x4_all}
    }
    \subfigure[{\Times{24}{24} MIMO}]{
      \includegraphics[width=3in]{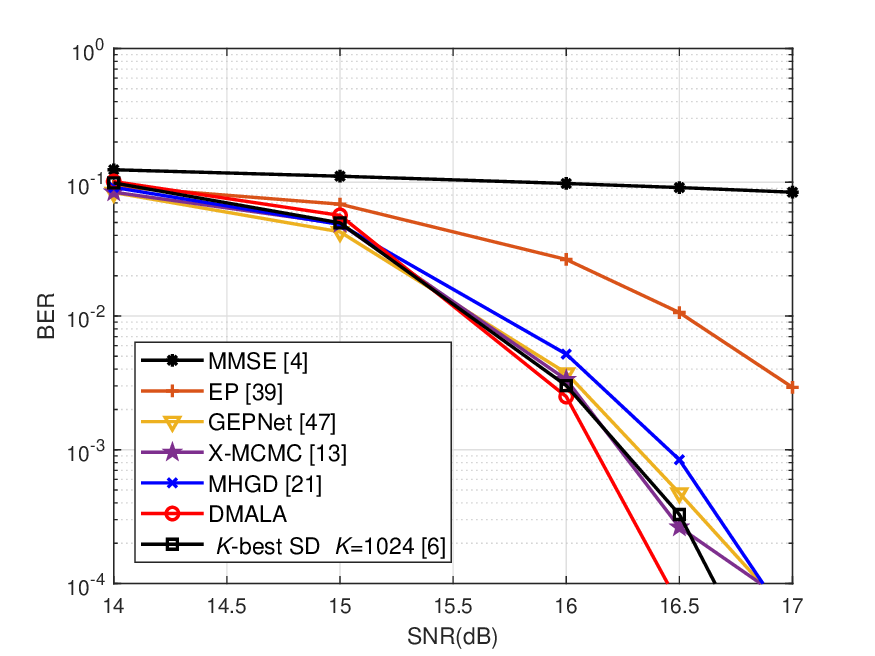}
      \label{fig:24x24_all}
    }
    \caption{{BER performance comparison in \Times{4}{4} and \Times{24}{24} MIMO systems employing 16-QAM modulation under Rayleigh fading channels. All MCMC-based detectors are configured with $N_{\rm p}=128$ and $N_{\rm p}=256$ parallel samplers for the \Times{4}{4} and \Times{24}{24} MIMO systems, respectively.}}  
    \label{fig:all}
  \end{figure}
}

In this subsection, we conduct performance and computational complexity comparisons with additional baselines, including the MMSE detector, and the Excited MCMC (X-MCMC) detector \cite{hedstromAchievingMAPPerformance2017}---an advanced Gibbs sampling-based MIMO detector. Furthermore, the performance of the graph neural network-enhanced EP detector (GEPNet) \cite{kosasih2022graph} is included to illustrate the comparison against state-of-the-art deep learning-based MIMO detection.\footnote{{We implement the GEPNet using the code released at \url{https://github.com/GNN-based-MIMO-Detection/GNN-based-MIMO-Detection}.}} We perform the comparison in two typical MIMO setups, including a small \Times{4}{4} MIMO system and a large \Times{24}{24} MIMO system. \figref{fig:all} presents the comparison results, showing that the proposed DMALA exhibits the most competitive performance across both system setups.

\CheckRmv{
	\begin{table}[t]
		\centering
		\caption{{Computational Complexity Comparison for the MIMO Detectors in \figref{fig:all}}}
    \setlength\tabcolsep{3pt} 
		{\begin{threeparttable}
			\begin{tabular}{ccc}
			\toprule
			{\multirow{2}{*}{\textbf{Algorithms}}} 
			& \multicolumn{2}{c}{\textbf{Number of FLOPs}}  \\
			\cmidrule{2-3}
			&\Times{4}{4} MIMO & \Times{24}{24} MIMO  \\
			\midrule
			MMSE \cite{yangFiftyYearsMIMO2015} & $6.60\times 10^2$ & $2.76\times10^4$ \\
			EP \cite{cespedesProbabilisticMIMOSymbol2018} & $3.14\times 10^4$ & $2.04\times 10^5$ \\  
			{$K$-best SD} \cite{guoAlgorithmImplementationKbest2006}  &  $1.54\times10^5$ ${\it \langle1.26\times10^7\rangle}$  & ${1.66\times 10^7}$ ${\it \langle5.40\times10^9\rangle}$  \\  
			GEPNet \cite{kosasih2022graph} & $4.90 \times 10^6$   & $1.71 \times 10^7$ \\
			X-MCMC \cite{hedstromAchievingMAPPerformance2017} & $9.20 \times 10^7\;(8.14\times 10^5)$  & $6.03\times 10^{9}\;(2.38\times 10^7)$ \\
			MHGD \cite{gowdaMetropolisHastingsRandomWalk2021}  & $2.63 \times 10^7\;(2.32\times 10^5)$  & $6.67\times 10^8\;(2.64\times 10^6)$ \\
			{DMALA} & ${1.45\times 10^7\;(1.47\times 10^5)}$ & ${1.53\times 10^8\;({6.53\times 10^5})}$ \\
			\bottomrule	
			\end{tabular}
			\label{tab:complexity}
	
			\begin{tablenotes}[para,flushleft]
				Note: ($\sim$) denotes the FLOPs count of a single MCMC sampler. $\langle\sim\rangle$ represents the number of comparison operations within the $K$-best SD detector, which is not included in the FLOPs count.
			\end{tablenotes}
		
		\end{threeparttable}}
	\end{table}
}
 
Subsequently, we illustrate the corresponding computational complexity of the compared detectors in \tabref{tab:complexity}, using the number of floating point operations (FLOPs) as the metric. This metric was profiled using the Python package for the performance application programming interface library \cite{papi}. The simulations were performed on a machine equipped with an Intel Xeon E5-4627 CPU at 2.6 GHz and 512 GB of memory. For MCMC-based detectors, the FLOPs count for both the whole detection process and one single sampler are presented. 
Furthermore, it is important to note that the considerable comparison operations\footnote{{The number of comparison operations within the $K$-best SD detector is separately recorded in \tabref{tab:complexity}, marked with $\langle\sim\rangle$.}} involved in the list administration of the $K$-best SD detector are not included in the FLOPs count, resulting in a complexity metric that favors the SD detector \cite{guoAlgorithmImplementationKbest2006,studer2010soft}. Based on \tabref{tab:complexity}, several observations are derived:
\begin{itemize}
  \item Compared to other MCMC-based detectors, including X-MCMC and MHGD, the proposed DMALA exhibits lower complexity in both the \Times{4}{4} and \Times{24}{24} MIMO systems. The advantage over X-MCMC can be attributed to the parallel update of all variables and the utilization of gradient information for acceleration. Additionally, the proposed method circumvents the high-complexity matrix inversion and line search involved in MHGD \cite{gowdaMetropolisHastingsRandomWalk2021}, thereby exhibiting higher computational efficiency. This superiority is particularly pronounced in the larger \Times{24}{24} MIMO setup, where DMALA achieves approximately a 77\% reduction in total complexity compared to MHGD.
  \item The overall computational complexity of the proposed DMALA is relatively higher than the $K$-best SD and GEPNet. Nonetheless, the proposed method benefits from parallel implementation, and the complexity of a single sampler is notably reduced. Particularly, in the \Times{24}{24} MIMO system, the complexity of a single DMALA sampler is significantly lower compared to both $K$-best SD and GEPNet. Therefore, by distributing the sampling process over these parallel samplers, the computation latency can be substantially reduced.
  \item The complexity increase from the \Times{4}{4} to the \Times{24}{24} MIMO setup in the proposed method is less pronounced when compared to the $K$-best SD. This favorable complexity scaling property of DMALA becomes evident especially when considering that a significant complexity component of the $K$-best SD is disregarded.
\end{itemize}

Taking a holistic view of the performance illustrated in \figref{fig:all} and the complexity presented in \tabref{tab:complexity}, our proposed detector showcases significant potential in striking a desirable balance between performance and complexity.}

\section{Conclusion}

In this study, we introduced a near-optimal MIMO detection scheme leveraging a novel gradient-based MCMC algorithm, designed for exact sampling within discrete spaces. This advancement enables the computation of highly accurate Bayesian estimates (soft decisions) through Monte Carlo techniques. We have rigorously proved the convergence of our proposed sampling algorithm, DMALA, and conducted a thorough analysis of its convergence rate. These theoretical underpinnings not only guarantee the performance of our proposed detector but also clearly delineate it from existing heuristic approaches. Our extensive numerical investigations empirically validate the near-optimality and resilience of the proposed method, affirming its superior performance compared to contemporary state-of-the-art MIMO detectors. Moreover, the proposed detector exhibits exceptional parallelization capabilities and scalability to the number of antennas, rendering it a viable and effective option for next-generation wireless communication systems.

{The current limitations of the proposed MCMC detector include increased computational overhead to achieve near-optimal performance and potential sensitivity to model mismatches. Future research should focus on improving the complexity scaling to accommodate ultra-large MIMO configurations and developing efficient hardware implementations for accelerated processing. Additionally, exploring a holistic approach that combines model-driven and data-driven methods would enhance the detector's adaptability to diverse channel conditions and system configurations.}

\appendices
\section{Proof of Lemma~\ref{th:property}}  \label{appendix1}
To avoid cluttered expressions, we consider the DMALA using the proposal in \eqref{eq:dis_prop_fac2}. Nonetheless, the proof also holds for the preconditioned version. 
We first show that the transition kernel of the Markov chain using DMALA is irreducible and aperiodic. For any two distinct states $\mathbf{x}$ and $\mathbf{x}^{\prime}$ from $\mathcal{A}^{N\times 1}$, the transition probability $P(\mathbf{x}^{(t+1)}=\mathbf{x}^{\prime}|\mathbf{x}^{(t)} = \mathbf{x})$ is given by $q(\mathbf{x}^{\prime}|\mathbf{x}) A (\mathbf{x}^{\prime}|\mathbf{x})$, as stated in \eqref{eq:trans_prob_1}, where the proposal probability $q(\mathbf{x}^{\prime}|\mathbf{x})$ and the acceptance probability $A(\mathbf{x}^{\prime}|\mathbf{x})$ are given in \eqref{eq:dis_prop_fac1} and \eqref{eq:acc_prob}, respectively. The proposal probability $q(\mathbf{x}^{\prime}|\mathbf{x})$ satisfies $0<q(\mathbf{x}^{\prime}|\mathbf{x})<1$ as long as the step size $\alpha>0$, given that $\left[\nabla f(\mathbf{x})\right]_n$ and $(x^{\prime}_n - x_{n})$ in \eqref{eq:exp_fac} are bounded and the proposal probability is normalized.
Moreover, based on \eqref{eq:f} and \eqref{eq:acc_prob}, the acceptance probability  can be written as
\CheckRmv{
  \begin{align}
    A(\mathbf{x}^{\prime}&|\mathbf{x}) \nonumber \\ 
      = &\min \left\{1, \exp\big(f(\mathbf{x}^{\prime}) - f(\mathbf{x})\big) \frac{q(\mathbf{x}|\mathbf{x}^{\prime})}{q(\mathbf{x}^{\prime}|\mathbf{x})}\right\} \nonumber \\
      = &\min \left\{1, \exp\left(\frac{\|\mathbf{y} - \mathbf{Hx}\|^2 - \|\mathbf{y} - \mathbf{Hx}^{\prime}\|^2}{\sigma^2}\right) \frac{q(\mathbf{x}|\mathbf{x}^{\prime})}{q(\mathbf{x}^{\prime}|\mathbf{x})}\right\},
      \label{eq:acc_prob_2}
  \end{align}
}
which remains nonzero as long as $\sigma^2 > 0$. This ensures that
\CheckRmv{
  \begin{equation}
    P\big(\mathbf{x}^{(t+1)}=\mathbf{x}^{\prime}|\mathbf{x}^{(t)}= \mathbf{x}\big) > 0,\; \forall \mathbf{x}, \mathbf{x}^{\prime}\in \mathcal{A}^{N\times1}, \mathbf{x}^{\prime}\neq \mathbf{x},   
  \end{equation}
}
thereby confirming the chain's irreducibility by definition \cite{norris1998markov}.

On the other hand, when $ \mathbf{x}^{\prime}=\mathbf{x}$, according to \eqref{eq:trans_prob_1} we have 
\CheckRmv{
  \begin{align}
    &P\big(\mathbf{x}^{(t+1)}=\mathbf{x}|\mathbf{x}^{(t)} = \mathbf{x}\big) \nonumber \\ &= q(\mathbf{x}|\mathbf{x}) + \sum_{\mathbf{z}\neq \mathbf{x}} q(\mathbf{z}|\mathbf{x})\big(1- A(\mathbf{z}|\mathbf{x})\big)  
     \geq q(\mathbf{x}|\mathbf{x}) > 0, 
    \label{eq:stay_prob}
  \end{align}
}
where the first equation is due to the fact that the chain stays in state $\mathbf{x}$ either when the proposal is $\mathbf{x}$, or when the proposal is $\mathbf{z}\neq\mathbf{x}$ but is rejected. 
The first inequality holds because $0<q(\mathbf{z}|\mathbf{x})<1$ based on the preceding analysis, and $0<A(\mathbf{z}|\mathbf{x})\leq 1$ according to \eqref{eq:acc_prob_2}.
The second inequality holds for the same reason as $q(\mathbf{x}^{\prime}|\mathbf{x})>0\; (\mathbf{x}^{\prime}\neq\mathbf{x})$. Leveraging \eqref{eq:stay_prob}, we deduce that
\CheckRmv{
  \begin{equation}
    \text{gcd}\left\{m: P\big(\mathbf{x}^{(t+m)}=\mathbf{x}|\mathbf{x}^{(t)}= \mathbf{x}\big) >0\right\} = 1
  \end{equation}
}
for any positive integer $m>0$, where ``gcd'' denotes the greatest common divisor.
This implies that the chain is aperiodic by definition \cite{norris1998markov}.

We then show that the chain induced by DMALA is reversible with respect to the target posterior distribution $\pi$, i.e., given two states $\mathbf{x},\mathbf{x}^{\prime} \in \mathcal{A}^{N\times1}$, the following equation (also called detailed balance condition \cite{bishopPatternRecognitionMachine}) holds:
\CheckRmv{
  \begin{equation}
    \pi(\mathbf{x}) P(\mathbf{x}^{\prime}|\mathbf{x}) = \pi(\mathbf{x}^{\prime}) P(\mathbf{x}|\mathbf{x}^{\prime}),
    \label{eq:detailed_balance}
  \end{equation}
}
which is a sufficient condition for guaranteeing that $\pi$ is a stationary distribution of the Markov chain. Since \eqref{eq:detailed_balance} naturally holds when $\mathbf{x}^{\prime} = \mathbf{x}$, we consider the case where $\mathbf{x}^{\prime} \neq \mathbf{x}$.
In this case, we have  
\CheckRmv{
  \begin{align}
    \pi(\mathbf{x}) P(\mathbf{x}^{\prime}|\mathbf{x}) 
    &= \pi(\mathbf{x})q(\mathbf{x}^{\prime}|\mathbf{x}) A (\mathbf{x}^{\prime}|\mathbf{x}) \nonumber \\
    &= \min \left\{\pi(\mathbf{x})q(\mathbf{x}^{\prime}|\mathbf{x}), {\pi(\mathbf{x}^{\prime})q(\mathbf{x}|\mathbf{x}^{\prime})}\right\}, 
  \end{align}
}
and by virtue of the same rationale,
\CheckRmv{
  \begin{align}
    \pi(\mathbf{x}^{\prime}) P(\mathbf{x}| \mathbf{x}^{\prime}) 
    &= \pi(\mathbf{x}^{\prime})q(\mathbf{x}|\mathbf{x}^{\prime}) A (\mathbf{x}|\mathbf{x}^{\prime}) \nonumber \\
    &= \min \left\{{\pi(\mathbf{x}^{\prime})q(\mathbf{x}|\mathbf{x}^{\prime})}, \pi(\mathbf{x})q(\mathbf{x}^{\prime}|\mathbf{x})\right\}. 
  \end{align}
}
Hence, we justify that \eqref{eq:detailed_balance} holds.
With this equation, we have 
\CheckRmv{
  \begin{align}
    \sum_{\mathbf{x}^{\prime} \in \mathcal{A}^{N\times1}}\pi(\mathbf{x}^{\prime})P(\mathbf{x}| \mathbf{x}^{\prime}) 
    &= \sum_{\mathbf{x}^{\prime} \in \mathcal{A}^{N\times1}}\pi(\mathbf{x})P(\mathbf{x}^{\prime}|\mathbf{x}) \nonumber \\
    &= \pi(\mathbf{x})\sum_{\mathbf{x}^{\prime} \in \mathcal{A}^{N\times1}}P(\mathbf{x}^{\prime}|\mathbf{x})
    = \pi(\mathbf{x}),
  \end{align}
}
completing the proof of $\pi$ being a stationary distribution of the chain using DMALA.
Furthermore, since the chain is irreducible, this stationary distribution is unique \cite[Corollary 1.17]{levinMarkovChainsMixing}, completing the proof of Lemma \ref{th:property}.

\section{{Proof of Lemma~\ref{lemma:ratio}}} \label{appendix_lemma2}
{The proposal distribution of DMALA can be expressed as
\CheckRmv{
  \begin{align}
    q&(\mathbf{x}^{\prime} | \mathbf{x}^{(t)}) \nonumber \\
    &= \frac{\exp \big(-\frac{1}{2\alpha} \|\mathbf{x}^{\prime} - \mathbf{x}^{(t)} - \frac{\alpha}{2}\nabla f(\mathbf{x}^{(t)}) \|^2 \big)}{\prod_{n=1}^{N} \sum_{x_n^{\prime} \in \mathcal{A}} \exp \big(-\frac{1}{2\alpha} |x_n^{\prime}-x_n^{(t)}-\frac{\alpha}{2}[\nabla f(\mathbf{x}^{(t)})]_n|^2 \big)}.
  \end{align}
}
Combining this equation with the expression of the target posterior distribution, we have
\CheckRmv{
  \begin{align}
    &\frac{q(\mathbf{x}^{\prime} | \mathbf{x}^{(t)})}{\pi(\mathbf{x}^{\prime})} \nonumber \\
    =\;& \frac{\exp \big(-\frac{1}{2\alpha} \|\mathbf{x}^{\prime} - \mathbf{x}^{(t)} - \frac{\alpha}{2}\nabla f(\mathbf{x}^{(t)}) \|^2 \big)}{\prod_{n=1}^{N} \sum_{x_n^{\prime} \in \mathcal{A}} \exp \big(-\frac{1}{2\alpha} |x_n^{\prime}-x_n^{(t)}-\frac{\alpha}{2}[\nabla f(\mathbf{x}^{(t)})]_n|^2 \big)} \nonumber \\
    &\cdot \frac{\sum_{\mathbf{s}\in \mathcal{A}^{N\times 1}} \exp\big( f(\mathbf{s})\big)}{\exp f(\mathbf{x}^{\prime})}  \nonumber\\
    \geq\;& \frac{\sum_{\mathbf{s}\in \mathcal{A}^{N\times 1}} \exp\big( f(\mathbf{s})\big)}{\prod_{n=1}^{N} \sum_{x_n^{\prime} \in \mathbb{Z}} \exp \big(-\frac{1}{2\alpha} |x_n^{\prime}-x_n^{(t)}-\frac{\alpha}{2}[\nabla f(\mathbf{x}^{(t)})]_n|^2 \big)} \nonumber \\
    &\cdot G(\mathbf{x}^{(t)},\mathbf{x}^{\prime}) \nonumber \\
    \geq\;& \frac{\sum_{\mathbf{s}\in \mathcal{A}^{N\times 1}} \exp\big( f(\mathbf{s})\big)}{\prod_{n=1}^{N} \sum_{x_n^{\prime} \in \mathbb{Z}} \exp \big(-\frac{1}{2\alpha} |x_n^{\prime}|^2 \big)} \cdot G(\mathbf{x}^{(t)},\mathbf{x}^{\prime}) \nonumber \\
    =\;&\xi\cdot G(\mathbf{x}^{(t)},\mathbf{x}^{\prime}),
  \end{align}
}
where the two inequalities follow the fact from the lattice theory \cite{micciancio2004worst}:
\CheckRmv{
  \begin{align}
    &\sum_{x_n^{\prime} \in \mathcal{A}}\exp \bigg(-\frac{1}{2\alpha} |x_n^{\prime}-x_n^{(t)}-\frac{\alpha}{2}[\nabla f(\mathbf{x}^{(t)})]_n|^2 \bigg) \nonumber\\
    \leq &\sum_{x_n^{\prime} \in \mathbb{Z}}\exp \bigg(-\frac{1}{2\alpha} |x_n^{\prime}-x_n^{(t)}-\frac{\alpha}{2}[\nabla f(\mathbf{x}^{(t)})]_n|^2 \bigg)  \nonumber \\
    \leq & \sum_{x_n^{\prime} \in \mathbb{Z}}\exp \bigg(-\frac{1}{2\alpha} |x_n^{\prime}|^2 \bigg), 
  \end{align}
}
completing the proof of Lemma~\ref{lemma:ratio}.}

\section{Derivation of \eqref{eq:llr_is}} \label{appendix2}
According to the IS-based Monte Carlo summation in \eqref{eq:is_mc}, 
by treating $p(b_k = \pm 1 | \mathbf{y})$ as the expectation $\mathbb{E}_{p(\mathbf{b}_{-k}|\mathbf{y})}[p(b_k=\pm 1|\mathbf{y}, \mathbf{b}_{-k})]$, these APPs can be approximated by
\CheckRmv{
  \begin{equation}
    \frac{1}{S}\sum_{s=1}^{S} p(b_{k}=\pm 1 | \mathbf{y}, \mathbf{b}_{-k}^{[s]}) \frac{p(\mathbf{b}_{-k}^{[s]} | \mathbf{y})}{\pi_{\rm a}(\mathbf{b}_{-k}^{[s]})},
  \end{equation}
}
where $\mathbf{b}_{-k}^{[s]}$ is obtained by dropping the $k$-th bit after demapping the sample $\mathbf{x}^{[s]}$ drawn from $\pi_{\rm a}$.
Note that herein we use the argument $\mathbf{b}$ for $\pi_{\rm a}$ to simplify the derivation of the LLR expression. Consequently, the posterior LLR of the $k$-th bit can be approximated as 
\CheckRmv{
  \begin{align}
    \hat{L}_k &= \log\frac{\sum_{s=1}^{S} p(b_{k}=+ 1 | \mathbf{y}, \mathbf{b}_{-k}^{[s]}) \frac{p(\mathbf{b}_{-k}^{[s]} | \mathbf{y})}{\pi_{\rm a}(\mathbf{b}_{-k}^{[s]})}}{\sum_{s=1}^{S} p(b_{k}=- 1 | \mathbf{y}, \mathbf{b}_{-k}^{[s]}) \frac{p(\mathbf{b}_{-k}^{[s]} | \mathbf{y})}{\pi_{\rm a}(\mathbf{b}_{-k}^{[s]})}} \nonumber \\
    &=\log\frac{\sum_{s=1}^{S} \pi_{\rm a}(b_{k}=+ 1 | \mathbf{b}_{-k}^{[s]}) \frac{p(b_k = +1, \mathbf{b}_{-k}^{[s]} | \mathbf{y})}{\pi_{\rm a}(b_k = +1, \mathbf{b}_{-k}^{[s]})}}{\sum_{s=1}^{S} \pi_{\rm a}(b_{k}=- 1 |  \mathbf{b}_{-k}^{[s]}) \frac{p(b_k = -1, \mathbf{b}_{-k}^{[s]} | \mathbf{y})}{\pi_{\rm a}(b_k = -1,\mathbf{b}_{-k}^{[s]})}},
    \label{eq:llr_is_append1}
  \end{align}
}
where the second equality follows from the Bayes rule $p(a_1, a_2) = p(a_2|a_1)p(a_1)$. For the conditional probability $\pi_{\rm a}(b_{k}=\pm 1 | \mathbf{b}_{-k}^{[s]})$, to avoid numerical instability, we consider a log-domain implementation by defining 
\CheckRmv{
  \begin{align}
    \gamma_{k}^{[s]} & =\log \frac{\pi_{\rm a}(b_{k}=+1 | \mathbf{b}_{-k}^{[s]})}{\pi_{\rm a}(b_{k}=-1 | \mathbf{b}_{-k}^{[s]})}=\log \frac{\pi_{\rm a}(b_{k}=+ 1 , \mathbf{b}_{-k}^{[s]})}{\pi_{\rm a}(b_{k}=- 1 , \mathbf{b}_{-k}^{[s]})} \nonumber \\
    & =\frac{1}{\tau}\left(f (\mathbf{x}_{+1}^{[s]})-f (\mathbf{x}_{-1}^{[s]})\right).
  \end{align}
}
Therefore, we have 
\CheckRmv{
  \begin{equation}
    \pi_{\rm a}(b_{k}=\pm 1 | \mathbf{b}_{-k}^{[s]}) = \frac{1}{1+\exp(\mp \gamma_k^{[s]})}. \label{eq:llr_is_append2}
  \end{equation}
}
Moreover, based on the definition of $\pi$ and $\pi_{\rm a}$, we have
\CheckRmv{
  \begin{equation}
    \frac{p(b_k = \pm 1, \mathbf{b}_{-k}^{[s]} | \mathbf{y})}{\pi_{\rm a}(b_k = \pm 1, \mathbf{b}_{-k}^{[s]})} \propto \frac{\pi(\mathbf{x}_{\pm 1}^{[s]})}{\pi(\mathbf{x}_{\pm 1}^{[s]})^{1/\tau}} = \exp\left(\frac{\tau-1}{\tau}f(\mathbf{x}_{\pm 1}^{[s]})\right).
    \label{eq:llr_is_append3}
  \end{equation}
}
Substituting \eqref{eq:llr_is_append2} and \eqref{eq:llr_is_append3} into \eqref{eq:llr_is_append1}, we derive \eqref{eq:llr_is}.



\ifCLASSOPTIONcaptionsoff
  \newpage
\fi




\end{document}